\documentclass{aa}  
\usepackage[colorlinks=true,linkcolor=blue,citecolor    = blue]{hyperref}
\usepackage{graphicx}
\usepackage{txfonts}
\usepackage{float}
\usepackage{rotating,tabularx}
\usepackage{multirow}
\usepackage{afterpage}
\usepackage{bigstrut}
\usepackage[table,xcdraw]{xcolor}
\usepackage[flushleft]{threeparttable }
\usepackage{amsmath}
\usepackage{enumitem}
\usepackage{amsfonts,color}
\usepackage{amssymb,float}
\usepackage[utf8]{inputenc}
\usepackage{color}
\usepackage{etoolbox}
\usepackage{appendix}
\usepackage{url}
\usepackage{dirtytalk}
\usepackage{float}
\usepackage{fontawesome}

\newcommand{\ra}[1]{\renewcommand{\arraystretch}{#1}}
\usepackage{booktabs}
\usepackage[graphicx]{realboxes}
\usepackage{adjustbox}
\usepackage{multirow}
\usepackage{rotating,tabularx}
\usepackage{afterpage}
\usepackage{tablefootnote}

\usepackage{tikz}

\usepackage{bigstrut}

\definecolor{lime}{HTML}{A6CE39}
\DeclareRobustCommand{\orcidicon}{%
    \begin{tikzpicture}
    \draw[lime, fill=lime] (0,0) 
    circle [radius=0.16] 
    node[white] {{\fontfamily{qag}\selectfont \tiny ID}};
    \draw[white, fill=white] (-0.0625,0.095) 
    circle [radius=0.007];
    \end{tikzpicture}
    \hspace{-2mm}
}

\foreach \x in {A, ..., Z}{%
    \expandafter\xdef\csname orcid\x\endcsname{\noexpand\href{https://orcid.org/\csname orcidauthor\x\endcsname}{\noexpand\orcidicon}}
}
\newcommand{\orcid}[1]{\href{https://orcid.org/#1}{\textcolor[HTML]{A6CE39}{\orcidicon}}}
\usepackage{graphicx}

\usepackage{txfonts}

\newcommand{\gaia}{\textit{Gaia}}

\begin{document}

\title{Variable stars in galactic globular Clusters I.\\
~The population of RR~Lyrae stars\thanks{Tables \ref{Tab:RR_Lyrae_GC}, \ref{Tab:constant_stars}, \ref{tab:cluster_parameters} and \ref{Tab:sources_sos} are available in electronic form at the CDS via anonymous ftp to cdsarc.cds.unistra.fr (130.79.128.5)
or via \url{https://cdsarc.cds.unistra.fr/cgi-bin/qcat?J/A+A/}} }

\titlerunning{RR~Lyrae stars in Galactic Globular Clusters}

   \author{Mauricio Cruz Reyes \inst{1}\orcid{0000-0003-2443-173X}
          \and Richard I. Anderson\inst{1}\orcid{0000-0001-8089-4419}\and  Lucas Johansson \inst{1}\orcid{0009-0001-1002-9603}
              \and Henryka Netzel\inst{1}\orcid{0000-0001-5608-0028} \and Zoé Medaric\inst{1}\orcid{0009-0006-1965-4241}      }

   \institute{Institute of Physics, \'Ecole Polytechnique F\'ed\'erale de Lausanne (EPFL), Observatoire de Sauverny, 1290 Versoix, Switzerland  \\
    \email{mauricio.cruzreyes@epfl.ch, richard.anderson@epfl.ch}   }

   \date{Received \today}

  \abstract
  {We present a comprehensive catalog of $2824$ RR Lyrae stars (RRLs) residing in $115$ Galactic globular clusters (GCs). Our catalog includes 1594 fundamental-mode (RRab), 824 first-overtone (RRc), and 28 double-mode (RRd) RRLs, as well as 378 RRLs of an unknown pulsation mode. We cross-matched 481\,349 RRLs reported in the third data release (DR3) of the ESA mission \gaia\ and the literature to 170 known GCs. Membership probabilities were computed as the products of a position and shape-dependent prior and a likelihood was computed using parallaxes, proper motions, and, where available, radial velocities from \gaia. Membership likelihoods of RRLs were computed by comparing cluster average parameters based on known member stars and the cross-matched RRLs. We determined empirical RRL instability strip (IS) boundaries based on our catalog and detected three new cluster RRLs inside this region via their excess \gaia\ $G-$band photometric uncertainties. 
  We find that $77\%$ of RRLs in GCs are included in the \gaia\ DR3 Specific Object Study, and $82\%$ were classified as RRLs by the \gaia\ DR3 classifier, with the majority of the missing sources being located at the crowded GC centers.
  Surprisingly, we find that $25\%$ of cluster member stars located within the empirical IS are not RRLs and appear to be non-variable. Additionally, we find that $80\%$ of RRab, $84\%$ of RRc, and $100\%$ of the RRd stars are located within theoretical IS boundaries predicted using MESA models with $Z=0.0003$, $M=0.7$ \(M_\odot\), and  $Y=0.290$. Unexpectedly, a higher $Y=0.357$ is required to fully match the location of RRc stars, and lower $Y=0.220$ is needed to match the location of RRab stars. Lastly, our catalog does not exhibit an Oosterhoff dichotomy, with at least 22 GCs located inside the Oosterhoff "gap," which is close to the mode of the distribution of mean RRL periods in GCs.
}


   \maketitle

\section{Introduction}
Globular clusters (GCs) are collections of thousands or even millions of stars with ages that range between $\sim 11$ to  $12.5$~Gyr \citep{2013ApJ...775..134V} and with a wide range of metallicities 
\citep[$\Delta \mathrm{[Fe/H]} = -2.3 $ dex,][]{harris2010new}.  They are particularly useful laboratories to understand the properties of pulsating stars because these variable members can be compared with (nearly) coeval non-variable stars, or other types of variables. Although multiple populations of stars 
\citep{2023MNRAS.520.1456L,2020MNRAS.491..515M} have been identified, GCs are relatively simple populations that can be used to test stellar models \citep{2008JPhCS.118a2010E}.

Globular clusters are known to host population\,II variable stars, notably including the RR~Lyrae stars (RRLs) studied here, as well as type-II Cepheids and Mira variables. In the 20th century, most studies dedicated to the identification of variable stars in clusters were conducted using photographic plates, which complicated the identification of variable sources with amplitudes smaller than $\sim 0.1$ mag. At that time, RRLs were the most commonly detected type of variables due to their high amplitudes and the relatively small baselines required for their characterization, and therefore they were known as cluster variables \citep{Hertzsprung1912,SHORE2003715}. 

\citet{1939PDDO....1..125S} carried out a pioneering work creating the first catalog of variable stars in clusters, which was built by collecting data from existing literature. Following the same approach, \citet[henceforth: C17]{2017EPJWC.15201021C} created the current most complete catalog of variable stars in GCs, which includes Eclipsing binaries (ECLs), SX Phoenicis, Type II Cepheids, slow variables, and RRLs. The catalog of C17 contains, in total, 5604 variables distributed in 122 GCs, out of which 2997 are RRLs and 114 are RRL candidates. At the time the catalog of C17 was created, the main criterion used to assess the membership of an RRL in a cluster relied on the measurement of its angular separation from the cluster center. It was commonly accepted that sources located closer to the center were more likely cluster members than those residing at larger angular separations. This approach was necessary because the vast majority of sources did not have a measured astrometry, and in some cases they did not have color information, making it impossible to identify their position in the color-magnitude diagram. Consequently, it is likely that some of them were incorrectly classified as cluster members. 

\citet{2022Univ....8..122B} conducted an extensive review of the current observational status of RRLs in GCs, mainly focusing on the photometric properties in both the optical and infrared domains. The review analyzes the light curves of these stars, their relationship with metallicity, and the morphology of the horizontal branch (HB) in the clusters that host them. Additionally, it offers a calibration of the period-luminosity relations for RRLs at multiple wavelengths.

Contributions that focus on obtaining the spectroscopy of individual clusters are crucial for determining the chemical composition of RRLs in GCs. The investigations led by  \citet{2018ApJ...864...57M,2019ApJ...881..104M} for NGC~3201 and NGC~5139~($\omega$~Cen) and the one led by \citet{2005ApJ...630L.145C} for NGC~6441 have contributed in this direction. These cluster-specific studies have shown that within a single cluster, there are multiple populations of RRLs. Consequently, in order to calibrate the period-luminosity relations for RRLs with maximum precision, it is not sufficient to use the average cluster metallicity, rather, individual measurements are necessary.

Recently, the third data release (DR3) of the ESA \gaia\ mission \citep{gaiamission,gaiaedr3} has delivered unprecedented astrometric measurements (positions, proper motions, and parallaxes) and multiband photometry of 1.8 billion objects. Additionally, \gaia's coordination unit seven (CU7) dedicated to variability has detected hundreds of thousands of RRLs and provided high-quality chromatic time-series photometry for them \citep{gaiadr3variability,gaiadr3rrl}. This wonderful dataset dataset allows for a homogeneous reassessment of the cluster membership of RRLs in GCs in unparalleled detail. 

Here, we aim to use the largest possible number of GCs and RRLs in conjunction with astrometry and photometry of from \gaia\ (DR3) to detect and classify RRLs in GCs.
The high source density near the centers of GCs introduces certain challenges for \gaia, notably when sources are no longer fully resolved. Hence, studying variable stars in GCs also allows one to test the limits of \gaia's performance, for example, in terms of completeness and contamination.

Our study takes advantage of the highly precise cluster parameters to evaluate the consistency of the theoretical instability strip (IS) boundaries with observations and explore whether all stars within the IS pulsate, a particularly interesting aspect considering previous findings indicate that around 30\% of stars within the IS boundaries for classical Cepheids do not exhibit pulsations \citep{2019MNRAS.489.3285N}. These analyses aim to improve our understanding of the models designed to explain stellar pulsations for RRLs. RRLs in GCs offer a unique laboratory for investigating the purity of the RRL IS because contamination is effectively minimized and because GC populations are much simpler than field populations.

This paper is the first one of a series focused on the study of variable stars within GCs. Section \ref{sec:description} provides a detailed description of the data employed in the membership analysis. Section \ref{sec:geometry} illustrates the method used to identify RRLs within clusters,  in particular Sects. \ref{sec:Prior}-\ref{sec:posterior} describe the approach utilized for the computation of the membership probabilities. 
The description of the results of the membership analysis, together with the detection of new RRLs, is presented in Sect. \ref{sec:M_analysis}. Our final sample of RRLs in GCs is presented in Sect. \ref{sec:Final}. The comparison with the theoretical models for the blue and red edges of the instability strip is shown in Sect. \ref{sec:models_is}, while Sect. \ref{sec:purity_is} discuss the non-variable stars located in this region.  Section \ref{sec:Oosterhoff} explains the Oosterhoff dichotomy and finally, Sect. \ref{sec:summary} summarizes the paper and presents our conclusions.

\section{Description of the data} \label{sec:description}

This section is divided into two parts: Section \ref{sec:clusters} describes the sample of clusters and their parameters, such as proper motion, parallax, metallicities and reddening estimates, whereas Sect. \ref{sec:RR_sample} describes the sample of RRLs used in our study.

\subsection{Cluster sample} \label{sec:clusters}
Using the astrometry of the \gaia\ early data release 3 \citep{gaiaedr3}, \citet[henceforth: VB21]{2021MNRAS.505.5978V}  identified the cluster member stars of 170 Milky Way GCs.  Our analysis relies on the clusters and the associated members identified by them. The VB21 dataset offers a membership probability for all stars located in regions near the cluster center, irrespective of the quality of their astrometry or photometry. This is extremely useful as it allows us to select the stars that best fit our needs. For example, when determining the central coordinates of a cluster, it is not recommended to remove sources with poor astrometry, as that would remove most of the sources located near the center, leading to an erroneous estimation of the central coordinates. On the other hand, to determine the cluster parameters, it is necessary to ensure that we use the highest quality astrometry, thus requiring strict quality cuts for both astrometry and photometry.  We updated the data of VB21 by crossmatching with \gaia\ DR3 \citep{2022arXiv220800211G}. All cluster parameters required for our study were recomputed here. A detailed description of the determination of cluster parameters will be presented separately (Cruz Reyes et al., in prep.). An abbreviated version is included in the Appendix \ref{sec:parameters}. Figure \ref{fig:histograms} displays the distributions of distances, metallicities, and reddening for the GCs in our sample.  

Cluster metallicities and reddening estimates were mainly taken from the catalog of cluster parameters by  \citet[henceforth: H10]{1996AJ....112.1487H,harris2010new}.  It is worth noting that H10 employs the metallicity scale provided by \citet{2009A&A...508..695C}. Seventeen clusters in the VB21 data set are not included in the H10 catalog, and therefore we searched for them in the literature.  For FSR~1758, the reddening is $\mathrm{E(B-V)} = 0.76 \pm 0.07$ \citep{2021AA...652A.158R}. The reddening and distance of FSR~1716 were estimated by \citet{2008AA...491..767B}  using isochrone fitting. However, we deem the resulting color excess unreliable because the derived distance $d =  0.8 \pm 0.1$~kpc is not consistent with the most recent estimate $d = 7.43 \pm 0.27$~kpc \citep[henceforth: BV21]{2021MNRAS.505.5957B}. We were unable to find $\mathrm{E(B-V)}$ for the remaining clusters.

Reddening coefficients were computed assuming an $R_V=3.3$ reddening law from \citet{1999PASP..111...63F} using \texttt{pysynphot} \citep{pysynphot} following \citet{Anderson2022} and the following parameters to mimic a typical RRL: $\mathrm{E(B-V)} = 0.5$ (mean of GCs, cf. Fig.\,\ref{fig:histograms}), $T_{\rm eff} =6800 K$, $ \log{g}=2.5$ and $\mathrm{[M/H] = -1.25}$. We thus obtained $R_{G}  = 2.855$, $R_{\rm Rp} = 2.047$, $R_{\rm Bp} = 3.576$, and $R_{W_{G}} = 1.867$ for the reddening-free Wesenheit magnitude. Furthermore, we obtained $\mathrm{E(Bp-Rp)} = 1.53 \times \mathrm{E(B-V)}$ and $\mathrm{E(Bp-G)}= 0.72\times \mathrm{E(B-V)}$. Given the wide range of color excess values in the GC sample, we recomputed the aforementioned values for $\mathrm{E(B-V)}=0.1$ and $1.0$ for comparison. This changes $R_{G}$, $R_{\rm Rp}$, and $R_{\rm Bp}$ by $\pm 5\%$ (higher $R$ for lower $\mathrm{E(B-V)}$), $R^W_G$ by $
\pm 2\%$, and $\mathrm{E(Bp-Rp)}$ by $\pm 3\%$. Restricting to clusters with $\mathrm{E(B-V)}\le 1.0$ limits possible biases in Wesenheit magnitudes to $< 0.02$\,mag, in $(Bp - Rp)_0<0.03$\,mag, and $< 0.05$\,mag in single-filter de-reddened magnitudes.

\begin{figure*}
    \centering
\includegraphics[scale= 0.8]{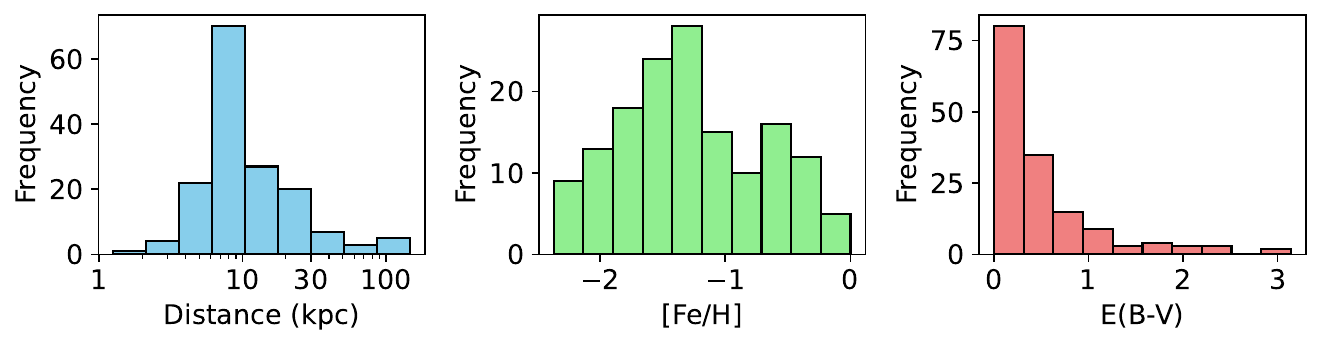}
    \caption{Distributions of distances, iron abundances, and reddening for our sample of GCs. } 
    \label{fig:histograms}
\end{figure*}

\subsection{The RR Lyrae sample}\label{sec:RR_sample}
For the convenience of readers and to enhance readability, this section contains all RRLs mentioned in this paper. These include all the RRLs detected by the \gaia\ collaboration, the RRLs in the C17 catalog crossmatched with \gaia\ DR3, and the RRLs detected in this paper.

\subsubsection{The Gaia sample}\label{sec:gaia_sample}
A key objective of the \gaia\ Collaboration, specifically of the CU7, is the identification and classification of all variable sources observed by \gaia. In the Third Data Release of \gaia\, the CU7 found 10.5 million variable sources \citep{gaiadr3variability}, which are reported in the table \texttt{vari\_classifier\_result}, of the \gaia\ archive\footnote{\url{https://gea.esac.esa.int/archive/}}. One million of those sources are active galactic nuclei and the rest are stars divided into 23 variability types, of which 297,778 are RRLs. 

The Specific Object Study (SOS) \citep[\texttt{vari\_rrlyrae}]{gaiadr3rrl} is designed to characterize the properties of Cepheids and RRLs, it contains 271,779 RRLs. The SOS analysis classifies each RRL according to their light curves, Fourier parameters, pulsation modes,  and it provides time series photometry for them.  We found 26\,202 RRLs in the classifier table that are not included in the SOS analysis, and 203 RRLs listed in the SOS analysis that are not present in the classifier list.  While the classifier aims to provide an analysis of the largest possible number of RRLs, the SOS analysis provides a more detailed analysis for bona fide RRLs. 

The catalog of RRLs of \gaia\ is among the most extensive ones, but it was not specifically designed to investigate regions with a high density of sources. \citet{2023A&A...674A..25H}  demonstrated the presence of spurious periodic signals related with the scanning law of \gaia. These types of signals could result in potential misclassifications in the \gaia\ catalog, specially for sources with close companions. This study explores \gaia's ability to observe and characterize RRLs in the extremely dense environments of GCs.

\subsubsection{RRLs from the literature}\label{sec:literature}
We complement our sample of RRLs using the catalog of \citet[henceforth: G23]{2022arXiv220701946G}, a compilation of 7\,841\,723 sources from 152 variable star catalogs cross-matched to sources reported in \gaia\ DR3. Among them, 393\,030 stars were identified as RRLs, out of which 183\,368 were reported neither by the classifier nor by the SOS analysis. The G23 catalog contains a boolean column labeled "selection" that allows the identification of the catalogs for which the classification is of higher quality. It is considered that the sources with "selection = False" are likely misclassified.

For our analysis, we combined the classifier, SOS, G23, and catalogs, to obtain a total of 481,349 unique RRLs. The cluster membership analysis, presented in Sect. \ref{sec:M_analysis} was applied only to sources that have astrometric solutions with 5 or 6 parameters in \gaia, in the range where systematic corrections to the parallax offset are defined by \citet[henceforth: L21]{lindegren2021gaia}. This  restricts our sample as follows: $6 < \mathrm{phot\_g\_mean\_mag < 21}$, $1.1 < \mathrm{nu\_eff\_used\_in\_astrometry} < 1.9$ (5-p sources), and $1.24 < \mathrm{pseudocolor} < 1.72$ (6-p sources). 

\subsubsection{C17 sample}\label{sec:clement}
As mentioned earlier, the C17 catalog contained the largest compilation of RRLs within clusters prior to our study. Section \ref{sec:Final} compares it with the results of our membership analysis. We crossmatched the sources in the C17 catalog with \gaia\ DR3 using a two arcsecond radius. We were able to identify 3015 sources out of the total 3111 candidate RRLs in the C17 catalog.  In some occasions multiple stars in the C17 catalog are associated with more than one source in \gaia, in those situations, we decided to prioritize the sources with the smallest angular separation.

\subsubsection{Additional RR Lyrae candidates}
Three RRLs were detected by analyzing the photometric uncertainties in the $G$ band of all stars located in the HB. These are presented in Sect. \ref{sec:HB}. 

\section{Identification of RR Lyrae in globular clusters} \label{sec:geometry} 
This section describes the method used to detect RRLs in GCs, which is based on hypothesis testing that is built by comparing the astrometric parameters of RRLs and clusters.  Section \ref{sec:Prior} describes the prior that we used for the membership analysis, Sect.~\ref{sec:likelihood} describes the likelihood and Sect.~\ref{sec:posterior} describes the posterior and the criteria that we use to consider an RRL as a likely cluster member.

\subsection{Cluster ellipticities}\label{sec:spatial}
More than one hundred years ago \citep{1917PNAS....3...96P}, it was discovered that the geometry of GCs is better described by ellipsoids than by spheres. We measure the ellipticities using the astrometry and photometry of the clusters members identified by VB21.

We start by modeling the two-dimensional geometric shape of clusters in RA and DEC using a principal component analysis (PCA). We decided not to incorporate higher-order principal components, as the structure of our dataset (RA and DEC) was effectively modeled by these components. The axes of each ellipse are given by the eigenvectors $\nu_{1},\nu_{2}$ of the PCA and their lengths by the eigenvalues $\lambda_{1}$ and $\lambda_{2}$.  The largest eigenvalue corresponds to the eigenvector pointing along the major axis of the ellipse ($a$) and the other eigenvector points in the direction of the minor axis ($b$). The position angle of the ellipse is measured with respect to its major-axis, with the following equation $\theta = \mathrm{arctan}( \nu_{a,y}/ \nu_{a,x})$, and the ellipticy is determined as $\epsilon = 1 - \sqrt{\lambda_{b}/\lambda_{a}}$.   Details regarding the application of this method to the VB21 dataset are described in the Appendix \ref{sec:parameters}. In this paper, we use the elliptical shapes only to estimate the prior of membership, and in a separate publication (Cruz \& Anderson, in prep.) we present a detailed discussion of the ellipticities. 

\subsection{Prior}\label{sec:Prior}
\begin{figure}
    \centering
    \includegraphics[scale=0.39]{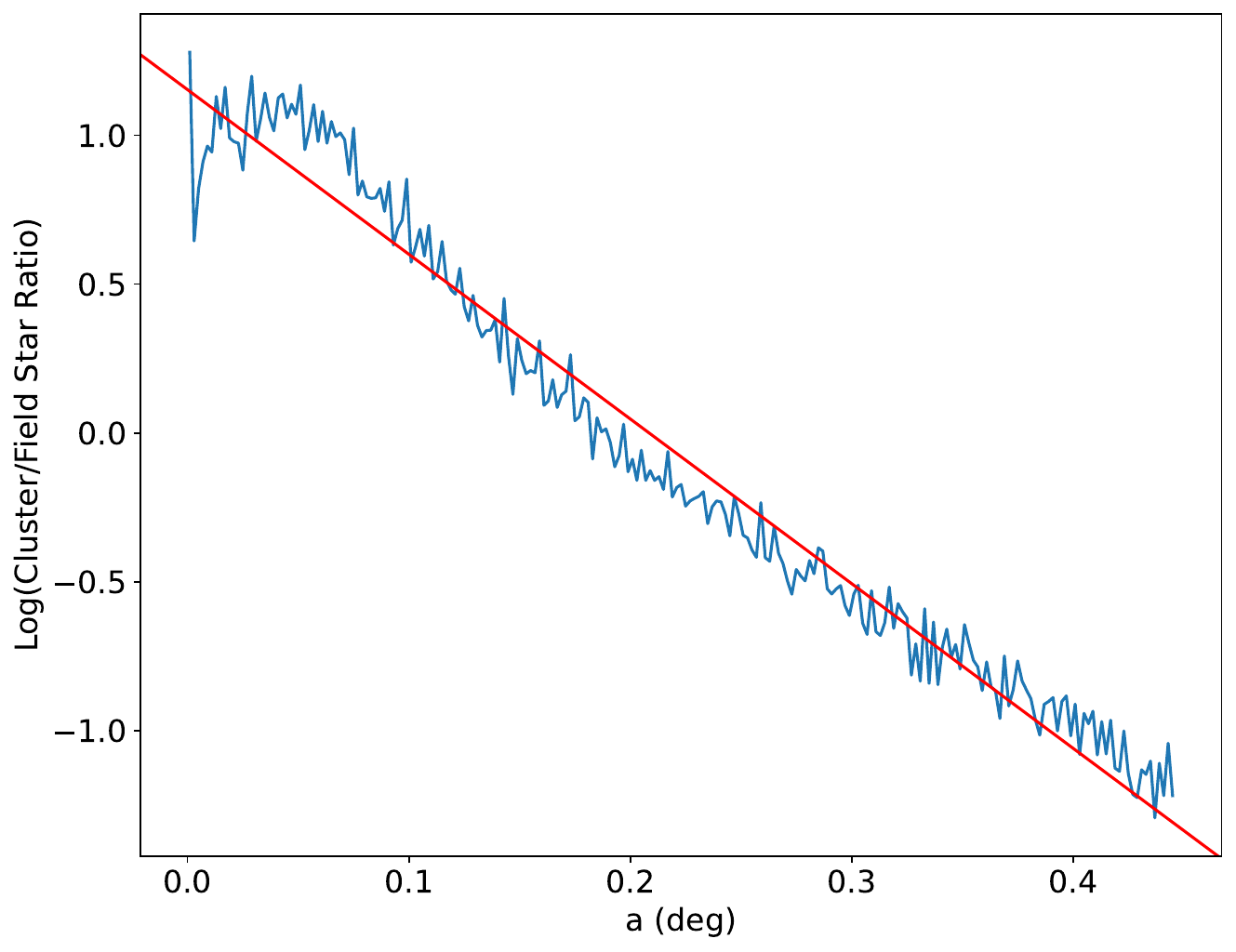}
    \caption{Base-10 logarithm of the number of cluster members divided by the number of field stars as a function of the angular distance from the cluster center for NGC~3201. The red line shows that this quantity decays approximately in a log-linear way. }
    \label{fig:exponential}
\end{figure}

Our goal is to derive a prior probability function that reflects our level of confidence regarding a star's membership in a cluster using only the star and cluster positions in the sky, and this information can be combined with the likelihood or considered separately. Since we expect most of the stars located near the core to be cluster members, the prior function should be unit-valued prior near the cluster center. Similarly, the prior should be approximately $1/10$ in regions where only one out of ten stars are clusters members.  To construct that function, we first defined a core ellipse centered on the cluster core. This core ellipse was aligned with the cluster members eigenvectors such that $66\%$ of stars are cluster members and the rest are background or foreground stars. Inside this core ellipse, we set the prior equal to one. We further defined a limiting ellipse such that only $10\%$ of stars are cluster members. A star falling on this limiting ellipse is assigned a prior probability of $1/10$. Since the ratio of cluster to field stars decays approximately exponentially (see Fig. \ref{fig:exponential}), we designed our prior function such that it also exponentially decreases outside the cluster core.

To determine the core and limiting ellipses we use the parameters derived with the PCA and the list of cluster members of VB21 (without astrometric or photometric quality cuts) and consider stars with membership probability greater than $50\%$ (as determined by VB21) as cluster members, and the rest as field stars.  We determine the ellipses in two different ways depending on the cluster. Clusters with more than 1000 members are run through a method in which we cut the cluster into different ellipse-slices, starting from a small ellipse around the cluster center and ending at the ellipse where its major axis is 1.2 times bigger than the maximum distance of a cluster star along that axis. To measure the core and limiting ellipses we analyze the ratio of cluster to field stars in each slice, as a function of the distance to the slice on the a-axis $a_{slice}$.  The points where the percentage of cluster stars is 66\% and 10\%, respectively (that is, the points where the cluster stars outnumber field stars by two, and field stars outnumber cluster stars by nine) are label as $a_{c}$ and $a_{lim}$. Using this definitions, the function that meets our requirements for the prior is
\begin{align}
f(a_{\mathrm{R}},b_{\mathrm{R}}) = \frac{\sqrt{n^{2}b_{\mathrm{R}}^{2} + a_{\mathrm{R}}^{2}} - a_{ \mathrm{c}}}{a_{\mathrm{lim}} - a_{\mathrm{c}}}, \quad P(A) = \min \left ( 10^{-f(a_{\mathrm{R}},b_{\mathrm{R}})} ,1 \right ),
\end{align}
where $a_{\mathrm{R}}$ and $b_{\mathrm{R}}$ are the coordinates of the star of interest in the space $(a,b)$. 

Clusters with less than 1000 cluster member stars in VB21 were excluded from this initial analysis, because given the low number of sources we cannot cut them into small enough slices without being susceptible to statistical noise. Instead, we developed a scaling relation based on the well populated clusters that we then apply to the poorly sampled ones.
For all clusters, we measure the standard deviation of the distribution of sources on the major axis $\sigma_{a}$. Subsequently, for each well populated cluster we estimate $a'_{c} = a_{c}/\sigma_{a}$,   $a'_{lim} = a_{lim}/\sigma_{a}$, and then we calculate their mean value $\bar{a'}_{c}$, $\bar{a'}_{lim}$. Finally, for the small clusters we scale their $\sigma_{a}$,   by $\bar{a'}_{c}$ and $\bar{a'}_{lim}$ to obtain their core and limiting ellipses.

\subsection{Likelihood}\label{sec:likelihood}
We approach the question of whether a RRL belongs to a cluster by treating it as a null hypothesis test, with the null hypothesis being that of membership.  We estimate the probability of membership using the Bayes theorem, which states that the posterior probability of
membership $P(A|B)$ is proportional to the product of the likelihood $P(B|A)$ and prior $P(A)$. To estimate the likelihood, we follow the methodology described in \cite{2013MNRAS4342238A} and \citet{2023A&A...672A..85C}. The likelihood is thus constructed as follows:
\begin{equation}
P(B|A) = 1 - p(c),     
\end{equation}
where $p(c)$, represents the level of confidence at which we can reject the null hypothesis of cluster membership. The $c$ quantity is defined as $c = \mathbf{x^T \Sigma^{-1} x}$ and is constructed using the parallax ($\varpi$) and proper motion ($\mu_{\alpha}^{*},\mu_{\delta}$). Radial velocity ($v_{r}$) measurements were additionally considered if a cluster has more than ten RV measurements and if the average RV of the RRLs was determined by the SOS analysis (table \texttt{vari\_rrlyrae}). The vector $\mathbf{x}$ is defined as 
\begin{align}
 x = (\varpi_{\mathrm{Cl}}-\varpi_{\mathrm{RR}},\mu_{\alpha, \mathrm{Cl}}^{*}-\mu_{\alpha, \mathrm{RR}}^{*},\mu_{\delta, \mathrm{Cl}}-\mu_{\delta, \mathrm{RR}},v_{r,\mathrm{Cl}} - v_{r,\mathrm{RR}}  ) \ ,    
\end{align}
and $\Sigma$ is the diagonal covariance matrix of the combined uncertainties of both RRL and cluster. 

The corrections to the parallax systematics of \gaia, provided by \citet{lindegren2021gaia} provide a good description of the parallax offset for faint sources $G>12$. However, additional adjustments are required for brighter sources \citep{2023arXiv230407158K,2022ApJ...938...36R}. 
If these residual corrections are not taken into account in the estimation of the likelihood, then they could lead to artificially low membership probabilities. In the case of RRL, 
\citet[henceforth: B21]{2021ApJ...909..200B} found that the residual parallax offset can be as big as $-22\,\mu$as, this is almost twice as large as the median parallax uncertainty for the clusters in our sample ($12\,\mu$as).  Therefore,  before computing the likelihood we added in quadrature $22\,\mu$as to the parallax uncertainty of all RRLs.

\subsection{Posterior}\label{sec:posterior}
We consider that a given variable star is a likely  cluster member if $P(A|B)>0.0027$, which corresponds to an overall difference below $3$ standard deviations with respect the cluster parameters, assuming $P(A) = 1$. It is crucial to emphasize that we cannot use the posterior to prove a matching pair, but only to reject it, if it is sufficiently small.

\subsection{Results of the membership analysis}\label{sec:M_analysis}
Using the method presented in the previous sections and the RRLs samples described in Sects. \ref{sec:gaia_sample} and \ref{sec:literature}, we identified 3620 RRLs distributed in 135 clusters, 2260 are included simultaneously in the SOS table and in the classifier, 197 are included only in the classifier and 1163 are present only in the G23 catalog, all of them have a likelihood greater than $0.0027$ and prior larger than $0.001$. The stars located outside the HB are removed from the sample in Sect. \ref{sec:Final}, as most of them are false cluster members, misclassifications, or spurious crossmatches.

\begin{figure}
    \centering
    \includegraphics[scale=0.38]{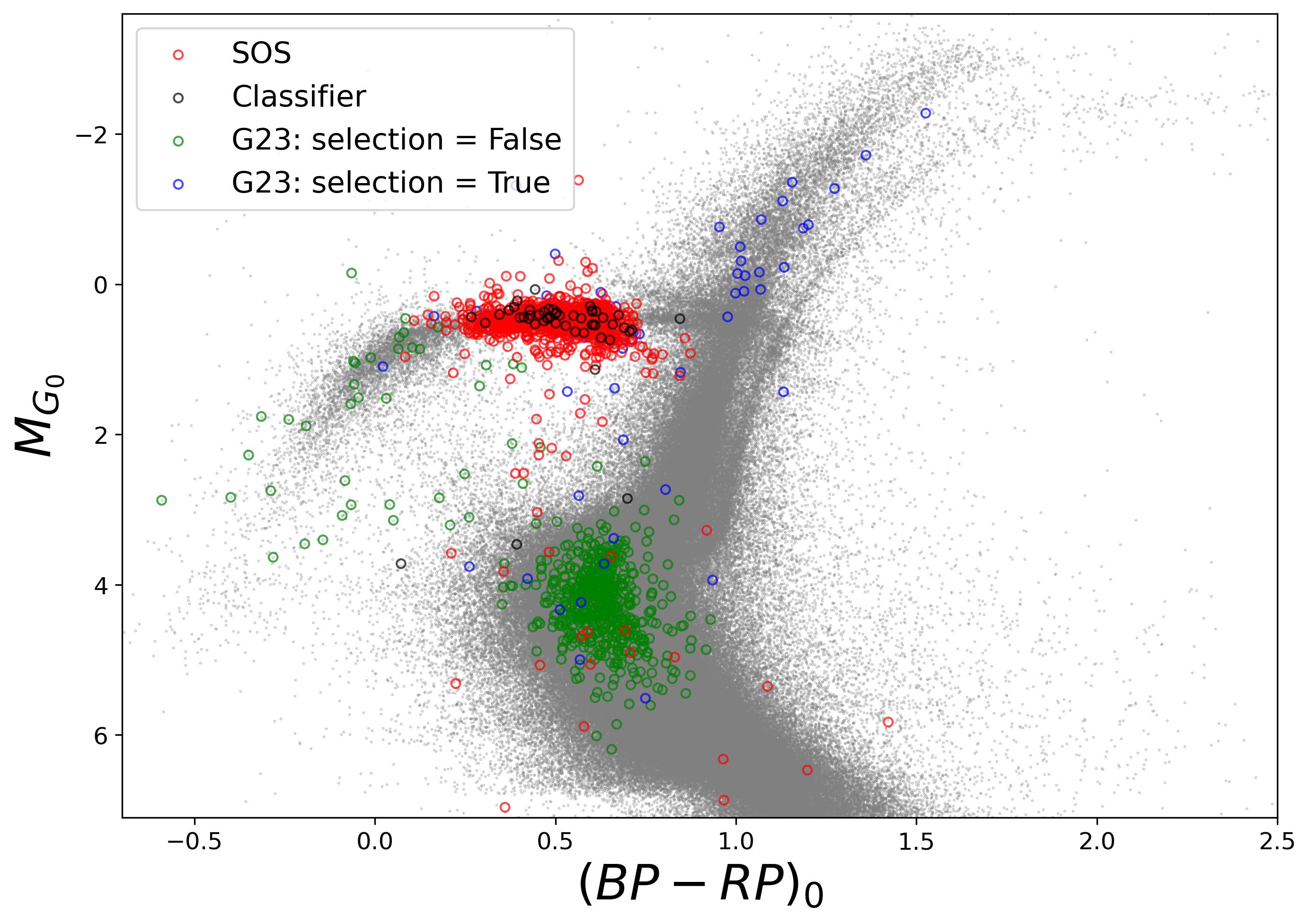}  \caption{Combined color-magnitude diagram of 75 GCs. Cluster members and RRLs meet the quality cuts described in Appendix \ref{app:IS}. Real RRLs in clusters are located in the HB, around $\mathrm{(BP - RP)}_{0} \sim 0.5$ mag and $M_{G_{0}} \sim 0.5 $ mag. Stars located outside the HB, are unlikely cluster members or not real RRLs. The G23 stars on the main sequence have the expected color for RRLs, but not the expected intrinsic absolute magnitude.  }
    \label{fig:is}
\end{figure}

Figure \ref{fig:is} displays the color-magnitude diagram for 75 GCs, with low reddening, and high quality cluster parameters, the exact criteria that were used to select them are explained in Appendix \ref{app:IS}. To plot the color-magnitude diagram we used the distances from BV21.  It can be clearly seen that a large number of sources are not located in the HB, which can be due to multiple reasons. For example, it is known that the precision in the astrometry of both RRL and clusters decreases as a function of distance, therefore, our method for computing membership probabilities loses the ability to reject false cluster members at large distances, increasing the contamination in our sample. 

Most of the sources in the G23 catalog with the selection flag set to False are located outside the HB. Nearly all of them were originally detected by \citet{2017AJ....153..204S}, but they were assigned a low probability of being RRLs; G23 classifies them as ECLs. Their location in the color-absolute magnitude diagram shows that the classification by G23 is likely correct, since the majority of them reside on the Main Sequence.

\begin{figure*}
    \centering
    \includegraphics[scale = 0.45]{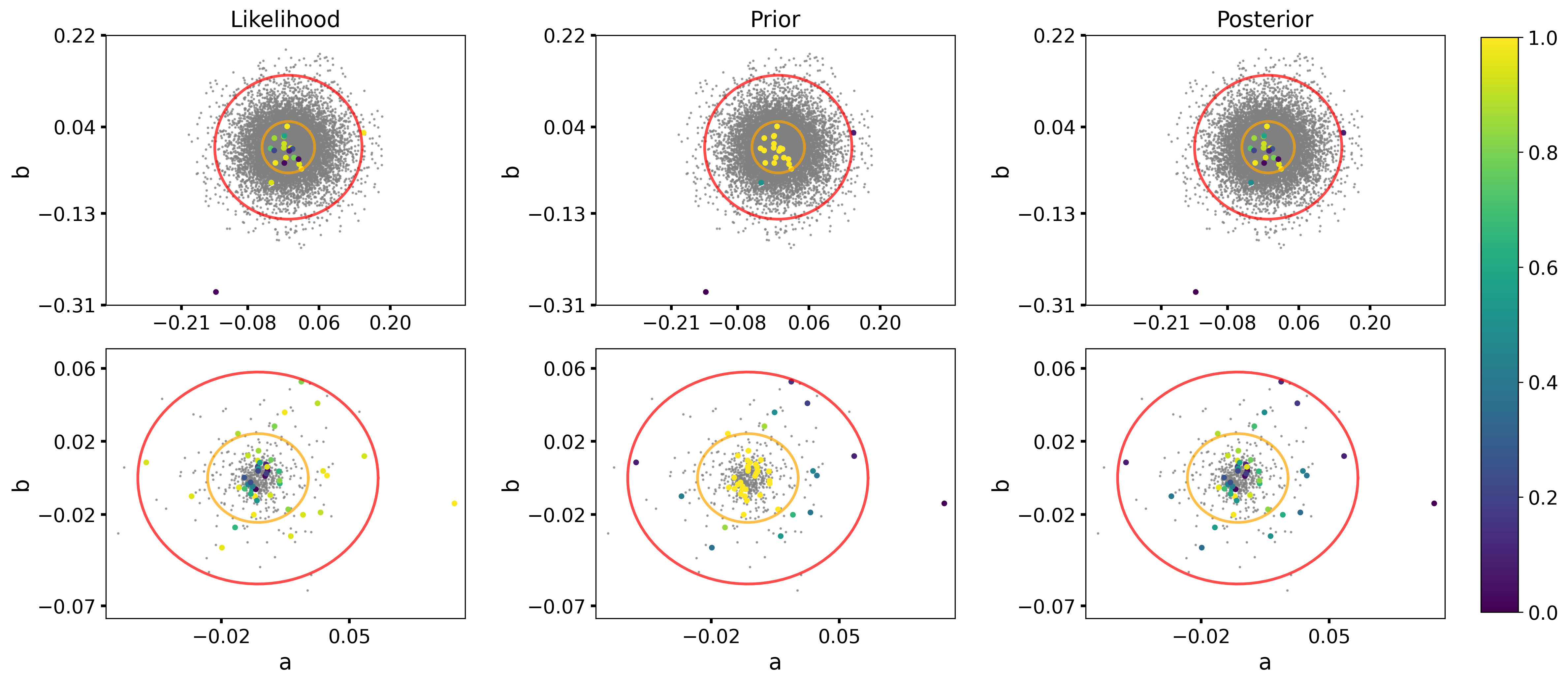}
    \caption{Cluster members of NGC~4833 and NGC~7006.  The color bar represents the membership probability assigned to the RRLs with our method. All stars were rotated to align the axes of the plot with the minor and major axes of the ellipses obtained with the PCA analysis. The orange ellipse represents the core ellipse and the red one the limiting ellipse. Both of them were obtained using the method presented in Sect. \ref{sec:Prior}  and relying on the sample of cluster members determined by VB21. At the limiting ellipse, only 10\% of all stars are expected to be cluster members. Field stars are not shown in the plot.  }
    \label{fig:probabilities}
\end{figure*}

Figure \ref{fig:probabilities} provides a graphical representation of the prior, likelihood, and posterior for the clusters NGC~4833 and NGC~7006. The first cluster contains more than 1000 stars in the VB21 dataset, and thus, the core and limiting ellipses were obtained using the first method described in Sect. \ref{sec:Prior}, while the ellipses of the second were obtained using the scaling relation.

\subsection{Undetected RRLs}\label{sec:HB}
\begin{figure}
    \centering
\includegraphics[scale=0.37]{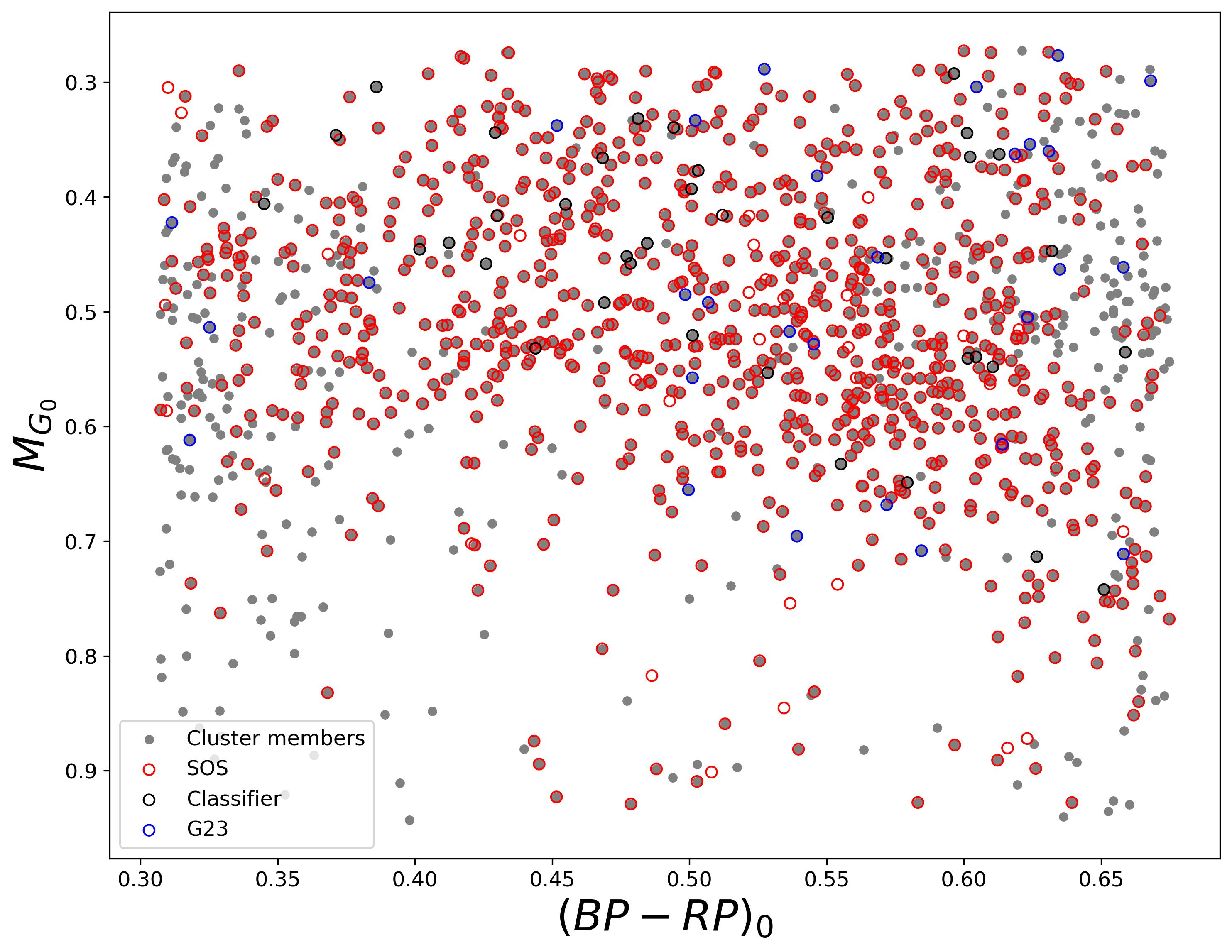} 
    \caption{Zoom in the region where most of the RRLs from \gaia\ are located  ($0.31 < \mathrm{(BP - RP)}_{0} < 0.67$ and $0.27 < M_{G_{0}} < 0.94$)  for 75 GCs. Not all cluster members in this region are detected as RRLs by any of the catalogs under our consideration. If a colored circle overlaps with a gray dot, it means that the particular cluster member was detected as an RRL.  }
    \label{fig:zoom_is}
\end{figure}

To find the region, where most of the RRLs are located, we decided to use only the stars belonging to the SOS and classifier catalogs. We further restrict our sample to the clusters that meet the quality criteria explained in Appendix \ref{app:IS}. The 5th and 95th percentile ranges of intrinsic color and absolute magnitude for RRLs in this sample are:
\begin{align}
    0.31 < \mathrm{(BP - RP)}_{0} < 0.67 \quad 0.27 < M_{G_{0}} < 0.94.
\end{align}\label{eq:o1}
Figure \ref{fig:zoom_is} shows this region. Not all cluster members located in this area have been reported to be RRLs, neither in \gaia, nor in G23, nor in C17. This either implies that some RRLs have not been detected by any of the catalogs under our consideration, or that not all stars located in the instability strip are RRLs.

To verify whether those cluster members located in the IS are unidentified RRL, we followed the same approach as \citet{2021A&A...648A..44M}, which consists in the identification of variable sources using the published weighted mean fluxes ($f_{G}$) and uncertainties ($\sigma_{f_{G}}$) from the \gaia\ catalog \citep{2021AA...649A...3R}. Assuming equally weighted measurements, the scatter in the light curves can be approximated as the weighted uncertainty in the flux ($\sigma_{f_{G}}$) multiplied by the number of observations. For constant stars, the scatter represents an estimate of the quality of the photometry that will vary depending on the magnitude of the source. However, for variable stars the scatters increases due to the astrophysical variability of the sources. Using this as an advantage it was found \citep{2021A&A...648A..44M} that a good unitless proxy for the amplitudes is given by

\begin{align}
    A(G) = \sqrt{N_{G}}\frac{\sigma_{f_{G}}}{f_{G}} \label{eq:amplitude}
\end{align}
where  $N_{G}$ is the number of observations in the $G$ band.   

\begin{figure}
    \centering
    \includegraphics[scale = 0.33]{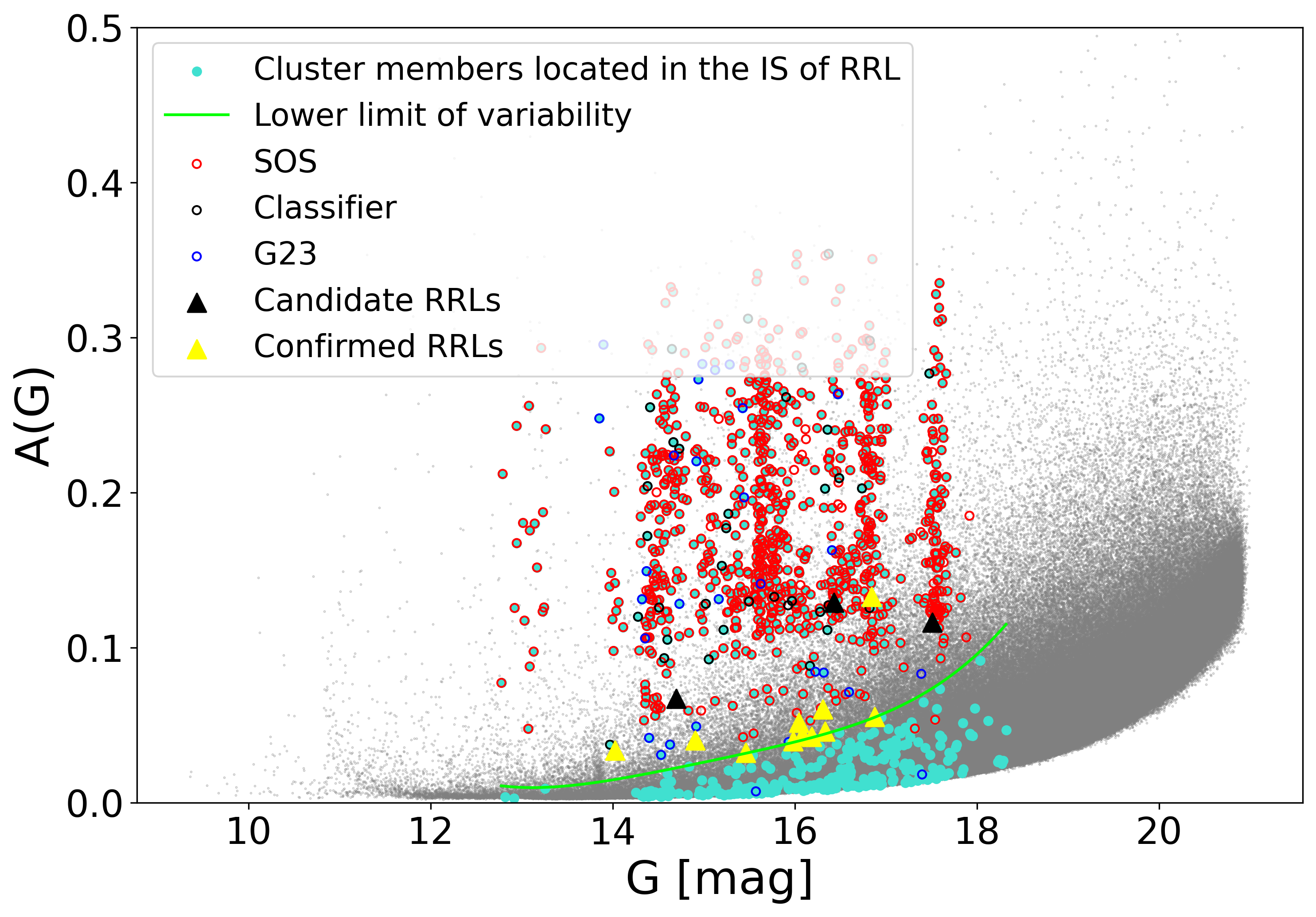}
    \caption{Unitless proxy amplitudes determined using Eq. (\ref{eq:amplitude}). Cluster members appear in a gray color, while the ones located in the region of the IS appear in turquoise. The yellow triangles denote confirmed RRLs that are not part of the \gaia\ catalogs presented in Sect. \ref{sec:description}. Meanwhile, the black triangles represent RRL candidates first reported here.}  
    \label{fig:amplitudes}
\end{figure}

We estimated the amplitudes of all cluster members using Eq. (\ref{eq:amplitude}), the results are shown in Fig. \ref{fig:amplitudes}. We delimited the region where constant stars are located by searching for the cluster members with the smallest amplitudes in bins of size 0.1 mag, and fitting a fourth order polynomial $P(G)$ to the results.  Additionally, we searched for the apparent magnitude ($G_{\mathrm{min}}$) of the RRLs from the SOS analysis with the smallest amplitude ($A_{\mathrm{min}}$) that was identified as a cluster member. We selected as potentially unidentified RRL, to the stars located in the IS with amplitudes  $A(G) >= \mathrm{limit} \times P(G) $, where  $\mathrm{limit} = A_{\mathrm{min}}/ P(G_{\mathrm{min}})$.

We found 14 potential RRLs, that are not included in the \gaia\ catalogs presented in Sect. \ref{sec:description}. However,  11 of them were previously identified as RRLs by other catalogs, therefore only three of the 14 are potentially new RRLs in clusters. Our results are summarized in Table \ref{tab:new_rr}. Unfortunately, there is no time-series photometry data available for the three candidates in \gaia\ DR3.

\section{Creating the final sample of RRLs in GCs}\label{sec:Final}
In this section we present the construction of the final sample of RRLs. As shown in Sect. \ref{sec:M_analysis}, a large number of RRLs reported in the literature are not located on the HB, indicating that they are not real RRLs or not actual cluster members.  As discussed in Sect. \ref{sec:gaia_sample}, some misclassifications may arise from spurious signals associated with the high density of sources in the clusters. Additionally, the likelihood presented in Sect. \ref{sec:geometry} may fail to reject RRLs with significant uncertainties in their astrometric parameters as cluster members. These stars were thus removed from the sample as described in Sect. \ref{sec:cleaning}. Meanwhile, in Sect. \ref{sec:comp} we compare our results with the C17 catalog. Our final set of RRLs in clusters is presented in Sect. \ref{sec:Final_final}.

\subsection{Removing false identifications}\label{sec:cleaning}
We selected all RRLs within three standard deviations of the median intrinsic color and absolute magnitude derived from the SOS and classifier samples. We have color excess measurements only for 154 clusters, using those values we identified 2422 RRLs as cluster members. For the remaining clusters, we individually inspected their color-magnitude diagram, in  BH~140 we identified five RRLs that are likely members and therefore they were added to our final sample, for the rest of the clusters we found that the RRLs are not located on the HB, and thus they are considered as false detections. In total, after including the newly detected RRLs, we were left with 2441 RRLs. 

\subsection{Comparison with the C17 sample}\label{sec:comp}
\begin{figure}
    \centering
    \includegraphics[scale= 0.26]{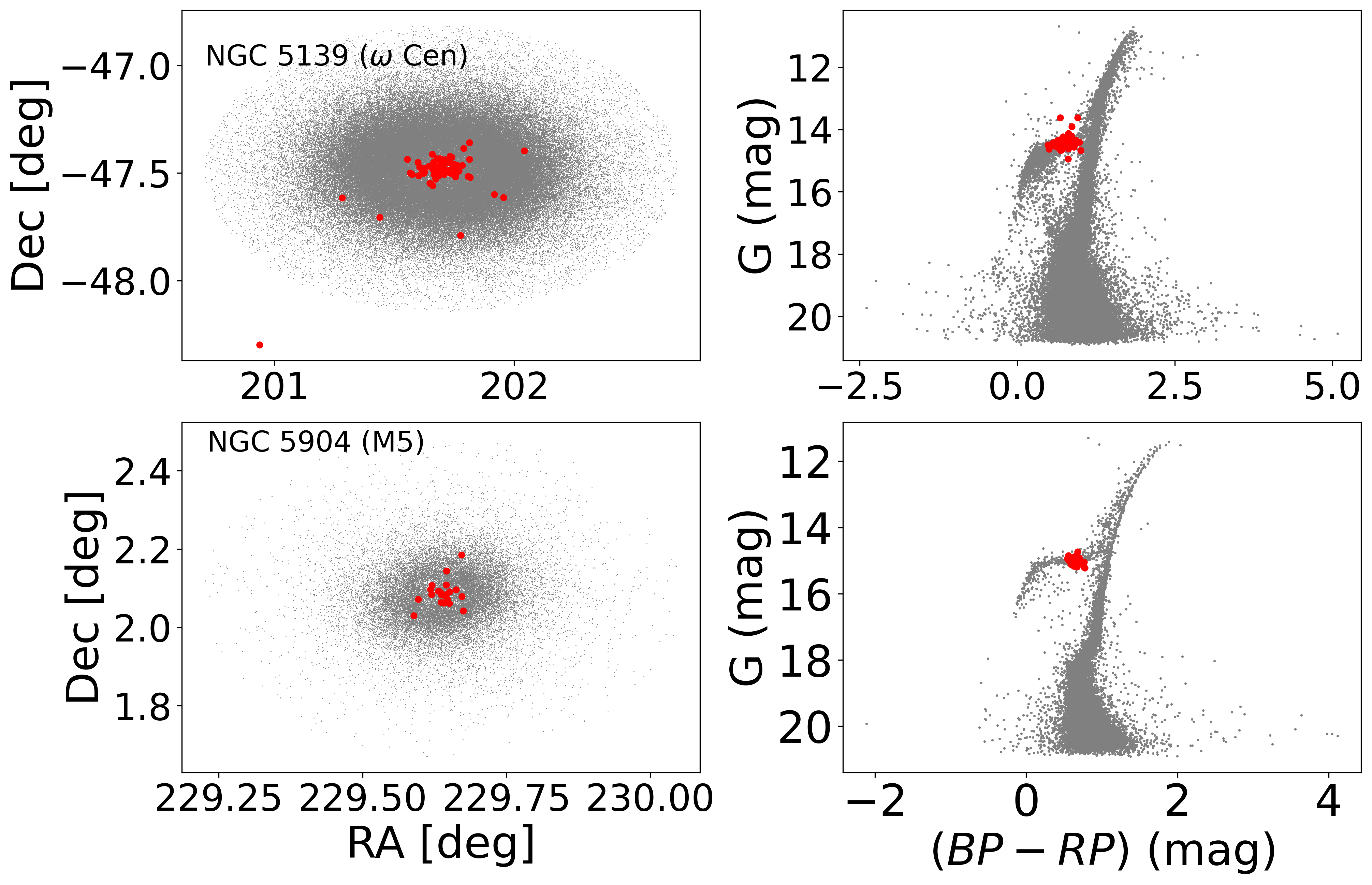}
    \caption{Position in the sky and color-magnitude diagram for  $\omega$~Cen and M5. Cluster members are represented by gray dots and the red dots indicate the RRLs that were not detected by the variable star analyses of the \gaia\ collaboration. We can observe that they are concentrated near the cluster centers.}
    \label{fig:crowding}
\end{figure}

From the 3015 RRLs in the C17 sample, 1806 are included in the sample presented in Sect. \ref{sec:cleaning}, 66 were rejected from the sample of RRLs in clusters with the method presented in the same section.   Nineteen of them have the selection flag equal to False, and 33 are identified with a different type of variability.  If we combine all the constraints, there are at least 112 unique RRLs in the C17 catalog that are unlikely cluster members or RRLs.  To remove the remaining outliers, we once again reject the three sigma outliers, following the approach presented at the beginning of Sect. \ref{sec:cleaning}. As a result, we identified 383 RRLs within clusters in the C17 catalog that were not initially detected in Sect.~\ref{sec:cleaning}. Most of them are located in the center of GCs. Twenty-three percent of the sources in this sample do not have parallax measurements in \gaia, hence they were not included in the membership analysis presented in Sect.~\ref{sec:M_analysis}. It is likely that our membership analysis failed to detect the remaining sources due to their poorer astrometry, as they exhibit an average ruwe value of 2.7, whereas those identified in Sect. \ref{sec:cleaning} have an average ruwe value of 1.5. Figure \ref{fig:crowding} illustrates the sources in this sample for the clusters NGC~5139~($\omega$~Cen) and NGC~5904~(M5). 

\subsection{Final sample}\label{sec:Final_final}
Our final set joins the results presented in Sects. \ref{sec:cleaning} and \ref{sec:comp}, it contains 2824 RRLs that reside in 115 GCs. The full sample of RRLs in clusters, including the rejected sources is presented in Table \ref{Tab:RR_Lyrae_GC}. This table contains, the \gaia\ DR3 source id, host cluster, prior, likelihood, and posterior for all sources. The sources that were detected in Sect. \ref{sec:comp} do not have a likelihood because they were not part of the analysis presented in Sect. \ref{sec:M_analysis}. In Sect. \ref{sec:M_analysis} we show that our initial sample contained 3620 RRLs in 135 clusters. To distinguish between the stars that pass our selection criteria from those that do not, we introduced a Boolean indicator "Final." If the indicator is set to True, it means they are considered cluster members, otherwise they were rejected because they are not located in the HB. This boolean column also allows the identification of sources from the C17 catalog that are not in our final sample. 

Thirty percent of the RRLs in C17 are not part of our final sample. However, this does not necessarily imply that these sources are not RRLs or cluster members. The C17 catalog is an inhomogeneous compilation of RRLs from various catalogs that includes RRLs detected using a variety of instruments covering optical (UBVI) to near-infrared ($J-$band) wavelengths. This inhomogeneity complicates detailed comparisons between the C17 sources and the sources cross-matched within the \gaia\ catalog, especially if the source was not identified as an RRL by \gaia. Positional cross-matches could be affected by blending, in particular if ground-based telescopes were used to identify RRLs in busy GCs. Since the majority of stars in a cluster are fainter than the HB, issues related to cross-matching would preferentially (but not exclusively) yield stars dimmer than the HB. On the other hand, blending will typically result in overly bright \gaia\ magnitudes and biased colors according to the type of blended object. Indeed, RRLs from C17 included in our final set have a mean value of ipd\_frac\_multi\_peak = 14, whereas this value is twice larger for the RRLs rejected in this work.

Table \ref{Tab:stats} shows the number of fundamental mode, first overtone, and double mode RRLs in our sample.  Within the entire set of RRLs identified in clusters, 2163 are identified by  \gaia's Specific Object Study (SOS), and 165 additional are identified by the classifier. The approximate completeness of these analyzes, relative to our catalog, is 77\% and 82\%, respectively.  From the sample presented in Sect. \ref{sec:M_analysis},  around $90\%$ of the stars from the SOS and classifier are within our final set. This contrasts sharply with the sources present only in the G23 sample, as only $4\%$ of the RRLs with the selection flag set to False are included, while $60\%$ of the sources with the selection flag set to True are in our final sample. This is interesting as it shows that the majority of stars identified as RRL by the \gaia\ collaboration are genuinely located in the HB, highlighting the robustness of the analysis even in areas with high star density.

\begin{table*}{}
  \centering
    \addtolength{\tabcolsep}{-1.7pt} 
  \caption{Results of the membership analysis.}
\begin{tabular}{llccccccccccc}\toprule  \\
\gaia\ DR3 source id & Cluster               & likelihood  & prior & posterior &  RV flag  & SOS  & classifier & G23  & C17  & New & Final\\   \midrule
4689637956899105792 & NGC 104 (47 Tuc)       &  0.3405     & 1     & 0.3405    & False    & True  & True       & True & True  & False & True\\
6045834869114826496 & NGC 6144               &  1.0000     & 1     & 1.0000    & False    & True  & True       & True & False & False & True\\
2342907756640334848 & NGC 288                &  0.9996     & 1     & 0.9996    & False    & True  & True       & True & False & False & True\\
2342908787434423040 & NGC 288                &  0.9897     & 1     & 0.9897    & False    & True  & True       & True & True  & False & True\\
... & ...   &  ... & ... & ... & ... &  ... & ... & ... & ... &  ...&  ... &  ...\\
              \bottomrule
\end{tabular}
\tablefoot{The complete version of this table is available at the CDS. RV flag is a Boolean column if True, it indicates that the SOS radial velocity was used to compute the likelihood. The columns SOS, classifier, G23, and C17 serve as boolean indicators, that specify if the source is included in the catalog at the top. The final column indicates the sources that are part of our final sample, as explained in \ref{sec:cleaning}. }
    \label{Tab:RR_Lyrae_GC}
\end{table*}

\begin{figure*}[ht]
    \centering
    \includegraphics[scale=0.53]{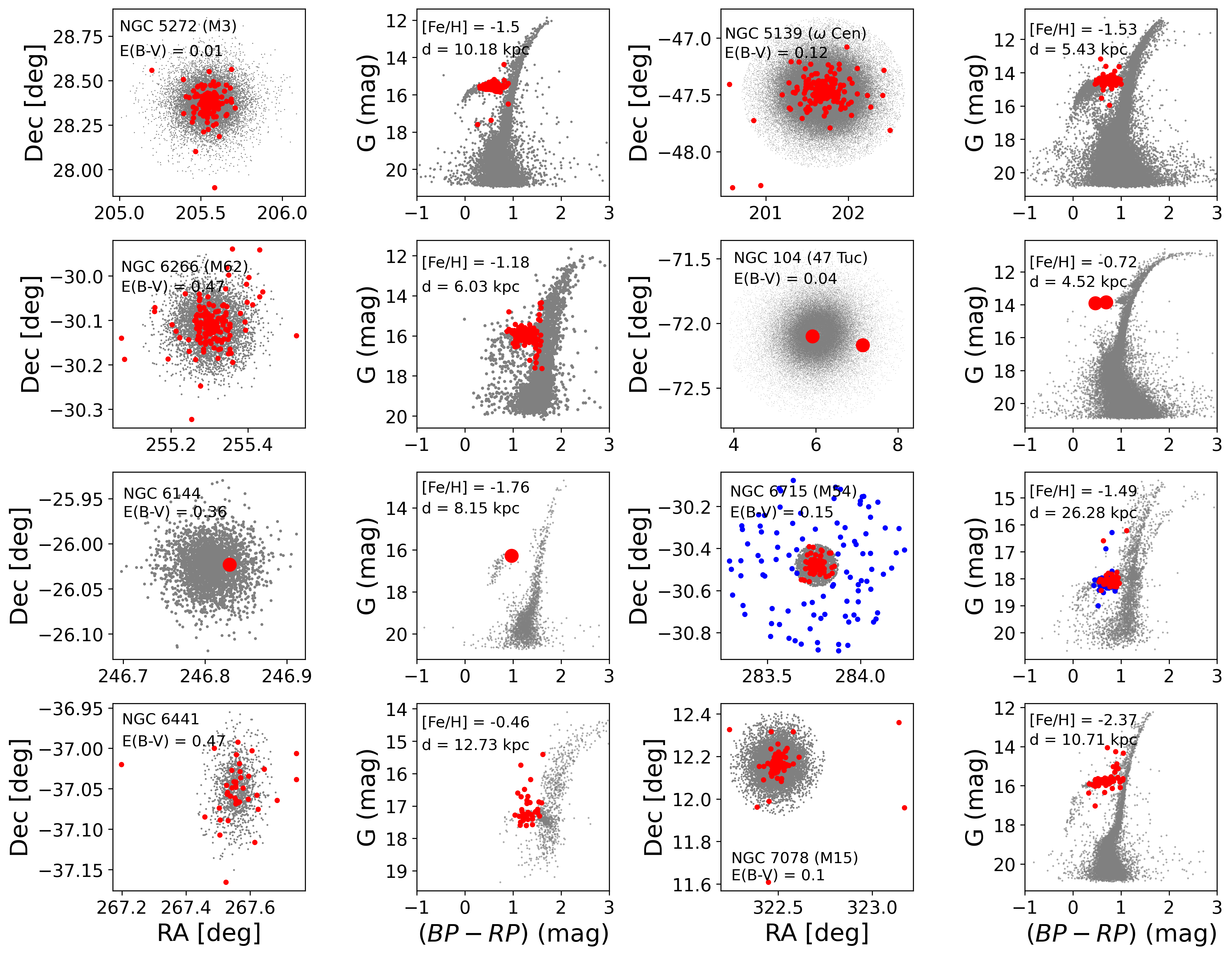}
    \caption{Position in the sky and color-magnitude diagram for some of the most interesting clusters in our sample. Cluster members appear in gray color and RRLs in red. For M54, the RRLs with prior smaller than 0.33 appear in blue, they are likely inside of the Sagittarius dwarf spheroidal galaxy but not inside the cluster. }      
    \label{fig:cluster_examples}
\end{figure*}

Figure \ref{fig:cluster_examples} shows the sky location and color-magnitude diagrams of eight particularly interesting clusters. Among them are the clusters with the highest and lowest number of RRLs in our sample, the particular case of NGC~6715~(M54), a cluster located within the Sagittarius dwarf spheroidal galaxy and finally two well populated clusters with particularly high and low metallicities.  The first three clusters, in Fig.  \ref{fig:cluster_examples} are NGC~5272~(M3), $\omega$~Cen and NGC~6266, they are the ones with the highest number of RRLs in our sample, with 236, 190, and 162 RRLs respectively. In contrast, some of the clusters with the lowest number of RRLs are 47~Tuc and NGC~6144. The number of RRLs in 47~Tuc has been highly debated in the literature. \citet{1980AJ.....85...36K} concluded that there are at least three stars of this type in the cluster. However, using radial velocities as membership constraint, \citet{1993PASP..105..294C} concluded that there is only one RRL member. Based on \gaia\ astrometry, we conclude that there are at least two RRL member stars. Given \gaia's limited spatial resolution, we cannot rule out the existence of more RRLs in the center of 47~Tuc. For NGC~6144, it appears that we have detected for the first time an RRL in this cluster. Interestingly, 47~Tuc and NGC~6144 have different iron abundances ($\Delta \mathrm{[Fe/H]}= 1.04$)  but both have almost zero stars near the HB. Both clusters exhibit distinctive features in their HB morphology. In 47~Tuc, the lack of stars in this region suggests that the stars may not have yet entered this evolutionary phase. Conversely, in NGC~6144, the stars within the HB are too blue to be RRL, as will be shown in Sect. \ref{sec:models_is} this might be related with a high helium content.

M54 presents a singular challenge. As noted by VB21 is not straightforward to separate the members of the dwarf galaxy from those of the cluster using their membership analysis.  The RRLs in our sample extend up to 0.42 degrees from the cluster center. At the cluster's distance of $26.28 \pm 0.33$~kpc (BV21), this corresponds to an extension of $192.5$ pc, which is clearly far too large given typical half-light radii of $\sim 5$ pc \citep{2019AARv..27....8G}. Using the prior,  $P(A)$, defined in Sect. \ref{sec:posterior}, we can remove the sources located at large angular separations from the cluster center. We found that the RRLs within the radius of the cluster obtained with the VB21 data have a prior greater than 0.33.  This value might seem unusually large, however, it is important to remember that the prior was estimated using the VB21 dataset, which, in this case, also contains sources from Sgr dSph. This contamination artificially increases the value of the prior. Figure \ref{fig:cluster_examples} displays the cluster sky location and the combined color-magnitude diagram for the Dwarf Galaxy and cluster.

Within all clusters containing RRLs, those with the highest number of RRLs have an iron abundance of $\mathrm{[Fe/H]} \sim -1.5$ (see Sect. \ref{sec:Oosterhoff}). In this context, M15 and NGC~6441 are particularly interesting. M3 is the cluster that contains RRLs with the lowest iron abundance in our entire set $\mathrm{[Fe/H]} = -2.37$, yet it hosts 133 RRLs. While, among the metal rich clusters, NGC~6441 ($\mathrm{[Fe/H]} = -0.46$) stands out as the one hosting the highest number of RRLs, reaching a total of 41.

\begingroup
\renewcommand{\arraystretch}{1.54}
\begin{table*}[!htp]
  \centering
  \caption{Number of RR Lyrae stars in globular clusters.}
\begin{tabular}{@{}lcccccccc}\toprule 
Type  &     $\mathrm{(BP - RP)}_{0} $  &      $M_{G_{0}} $   &  Number SOS & Number classifier &  Number G23 & Number new  &  Number C17 & Total\\ 
      &            (mag)  &       (mag)  &   & &   &  \\  \midrule
RRab     & $0.58^{+0.13}_{-0.15}$ &  $0.54^{+0.31}_{-0.31}$    &    1403    &        0           &   153 &  0  &  38  & 1594 \\
RRc      & $0.42^{+0.16}_{-0.16}$ &  $0.48^{+0.30}_{-0.20}$    &    736    &         0           &   67  &  0 & 21  &  824  \\
RRd      & $0.46^{+0.08}_{-0.08}$ &  $0.49^{+0.05}_{-0.22}$    &    25     &         0           &  3    & 0  & 0  &  28   \\
RR (Not classified) &           - &           -                &    0      &       165           &  192  &  14 & 7 &  378  \\
\midrule
\textbf{All RRLs} & $0.54^{+0.16}_{-0.23}$ &$ 0.51^{+0.35}_{-0.28}$    &  2164   &  165   &   415    &  14 &   66 & 2824 \\
\bottomrule
\end{tabular}
\tablefoot{The second and third column show the median intrinsic color and absolute magnitude of RRLs and the uncertainties represent the 5th and 95th percentiles, they were estimated using only the stars present in the \gaia\ SOS sample with $\mathrm{E(B-V)}$ measurements in our sample. The unique number of RRLs in each catalog can be found in the remaining columns.  The \gaia\ DR3 source id for those stars can be found in Table \ref{Tab:RR_Lyrae_GC}.} \\
    \label{Tab:stats}
\end{table*}

\endgroup

\section{Models versus observations}\label{sec:models_is}
We determined the blue and the red edges of the instability strip for pulsations in the fundamental mode (RRab) and the first-overtone (RRc) RRLs using MESA-RSP \citep{mesa_rsp} code with MESA version r23.05.1. We calculated a grid of models for $M=0.7 \,{\rm M}_\odot$, two metal abundances $Z=0.0003$, $0.0001$ and four helium abundances
$Y=0.220$, $0.245, 0.290, 0.357$ (for details how $Y=0.220$, $Y=0.290$ and $Y=0.357$ were chosen see further in the text). 
The models were computed in the luminosity range $\log L/$L$_\odot \in \langle 1.5, 1.8 \rangle$ with a step of 0.05\,dex, and effective temperature $T_{\rm eff} \in \langle 5600,8100 \rangle$\,K with a step of 50\,K. The mixing length parameter, $\alpha_{\rm MLT}$, was set to 1.5 (the rest of the convective parameters correspond to Set B in table~4 in \citealt{mesa_rsp}). The OPAL opacities were used \citep{iglesias1996}. The grid of models was transformed into {\it Gaia} photometric system using the PARSEC database of bolometric correction \citep[using EDR3 band definitions as in][]{ybc}. The linear growth rates for the fundamental and the first-overtone modes were interpolated to obtain the blue and red edges. The models are presented in Table \ref{tab:models}.

\begin{table*}{}
 \centering
\caption{Theoretical blue and red edges of the instability strip for RRLs in the ($G$,$BP-RP$) plane for pulsations in fundamental mode (RRab) and first overtone (RRc). \label{tab:models}}
\begin{tabular}{ccccc}\toprule
\multicolumn{1}{c}{Type} &  \multirow{1}{*}{Blue edge} &  \multirow{1}{*}{Red edge} &  \multirow{1}{*}{Parameters}   \\
\midrule

RRab &  $-17.02(BP - RP - 0.5) -1.64$  &  $-9.25(BP - RP - 0.5) + 2.00$  &  $Z=0.0003$, $M=0.7$ \(M_\odot\), $Y=0.220$   \\
RRc & $-11.24(BP - RP - 0.5) - 1.50$ & $-15.65(BP - RP - 0.5) + 1.84$ &  $Z=0.0003$, $M=0.7$ \(M_\odot\), $Y=0.220$    \\
\midrule
RRab & $-14.45(BP - RP - 0.5) - 1.73$  &  $-9.22(BP - RP - 0.5) + 1.72$ &  $Z=0.0003$, $M=0.7$ \(M_\odot\), $Y=0.245$     \\
RRc &  $-14.24(BP - RP - 0.5) - 2.41$  & $-14.02(BP - RP - 0.5) + 1.39$ &  $Z=0.0003$, $M=0.7$ \(M_\odot\), $Y=0.245$    \\
\midrule

RRab & $-15.16(BP - RP - 0.5) - 1.78$ &  $-9.42(BP - RP - 0.5) + 1.75$ &  $Z=0.0001$, $M=0.7$ \(M_\odot\), $Y=0.245$ \\
RRc & $-11.28(BP - RP - 0.5) - 1.76$  &  $-15.65(BP - RP - 0.5) + 1.46$ &  $Z=0.0001$, $M=0.7$ \(M_\odot\), $Y=0.245$     \\

\midrule
RRab & $-11.31(BP - RP - 0.5) - 1.40$ &  $-9.33(BP - RP - 0.5) + 1.60$ &  $Z=0.0003$, $M=0.7$ \(M_\odot\), $Y=0.290$   \\
RRc & $-13.33(BP - RP - 0.5) - 2.53$  & $-10.00(BP - RP - 0.5) + 0.87$ &  $Z=0.0003$, $M=0.7$ \(M_\odot\), $Y=0.290$    \\

\midrule
RRab &  $-14.04(BP - RP - 0.5) -2.13$  &  $-8.13(BP - RP - 0.5) + 1.21$  &  $Z=0.0003$, $M=0.7$ \(M_\odot\), $Y=0.357$   \\
RRc & $-24.61(BP - RP - 0.5) - 5.66$ & $-14.52(BP - RP - 0.5) + 0.57$ &  $Z=0.0003$, $M=0.7$ \(M_\odot\), $Y=0.357$    \\

\bottomrule
\end{tabular}
\end{table*}

\begin{figure}
    \centering
    \includegraphics[scale= 0.7]{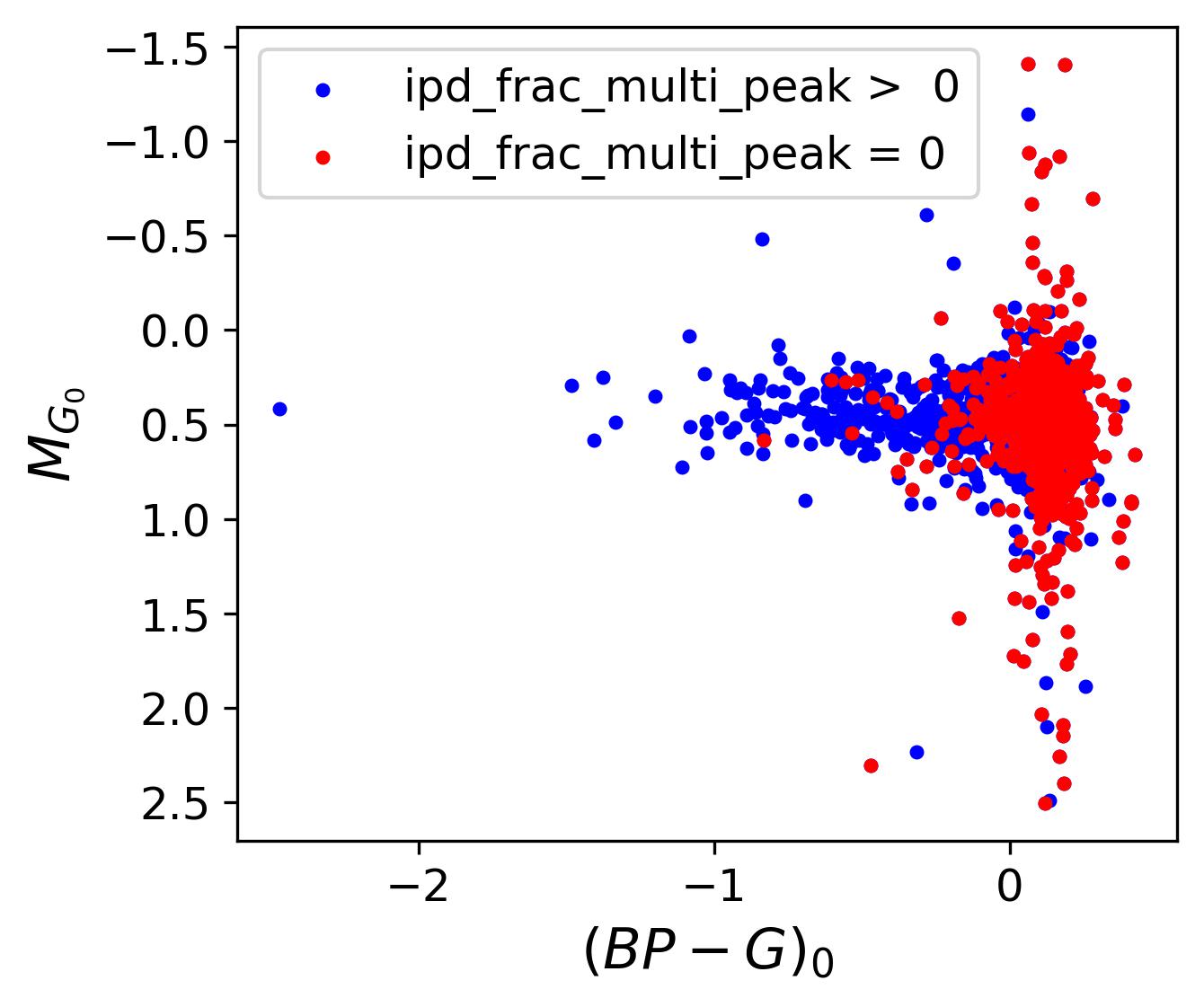}
    \caption{Color-magnitude diagram for the RRLs in our final sample. It can be seen that multiple sources have an unexpected 
$\mathrm{(BP - G)}_{0}$ color, most of them have a nonzero value of ipd\_peak\_multi\_frac, indicating that their photometry is likely affected by nearby sources.  }
    \label{fig:cmd_gaia}
\end{figure}

To compare the models with the observations, we considered only RRLs with mean magnitudes determined by the \gaia\ DR3 SOS \citep{gaiadr3rrl}. As a result, any sources not present in the SOS sample were not considered in the comparison of IS boundaries.  Figure \ref{fig:cmd_gaia} shows that there is a large number of sources with $\mathrm{(BP - G)}_{0} < 0$, the anomalous color of these stars is likely related with photometric blending, as for them the mean value of ipd\_frac\_multi\_peak is equal to $24$, while for sources with $\mathrm{(BP - G)}_{0} > 0$ the mean ipd\_frac\_multi\_peak is equal to three.  To avoid issues with blending and to compare directly with the models, we further restricted our sample of RRLs to those with ipd\_frac\_multi\_peak $= 0$.

\begin{figure*}
    \centering
    \includegraphics[scale = 0.57]{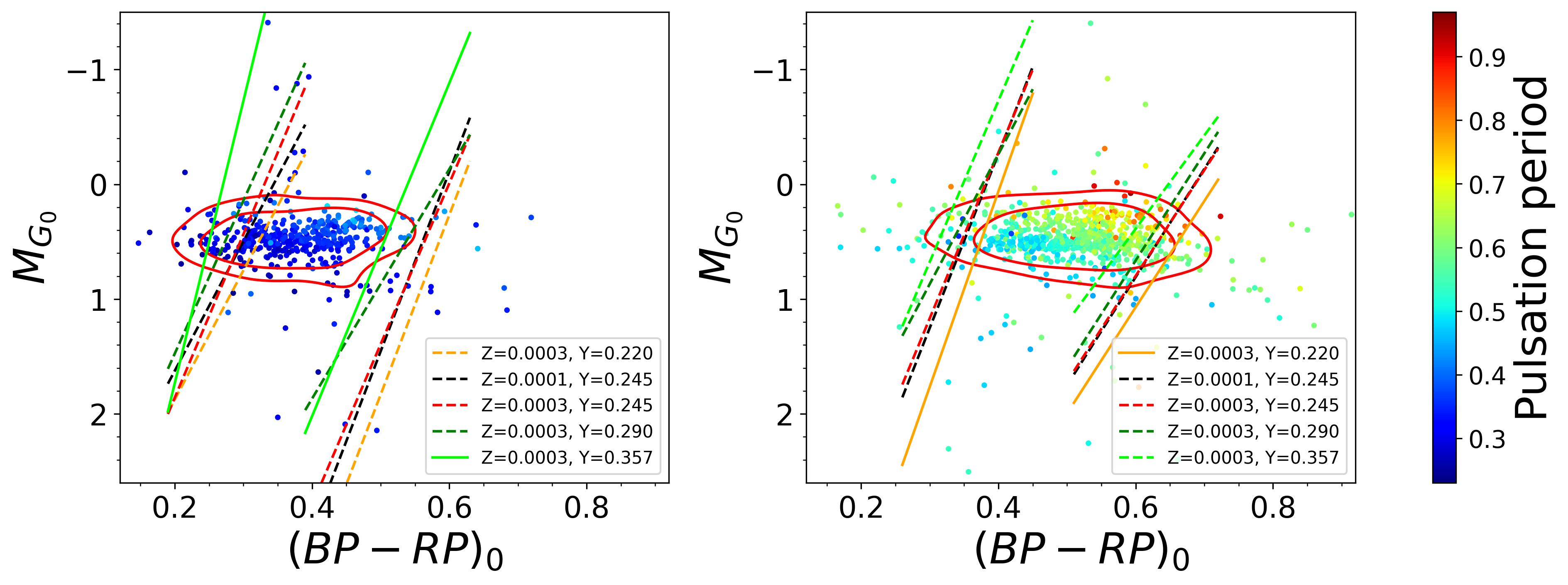}
    \caption{RRLs in the color-magnitude diagram color coded with their pulsation period. The first panel shows the RRc stars and the second panel shows the RRab. The color bar represent the pulsation period of the stars and the contours the 68th and 84th percentiles of density of RRL. All models assume $M=0.7$ \(M_\odot\). The dotted lines are the theoretical models for the blue and red edge of the instabillity strip of RRc and RRab stars respectively. The solid lines highlight the models that best fit the observations.}
    \label{fig:models_vs_obs}
\end{figure*}

In Fig. \ref{fig:models_vs_obs}, we show a comparison of the observations with the models. We find that models with a $Y=0.245$ fit relatively well the density contours for RRab stars. However, the same models are too red to fit the RRc stars, which are better described by models featuring a much higher helium abundance of $Y=0.357$. 
NGC 2419 and NGC 6266 (M62) contain simultaneously RRab and RRc stars beyond the IS boundaries predicted by the models with $Y=0.245$. However, NGC 2419 is situated at a relatively large distance of 88.47 kpc, and NGC 6266 is located in a region with significant differential reddening \citep{2010AJ....140.1766C}. Only one cluster (NGC 6362) contains RRc stars outside the boundaries matching RRab types, and three clusters (NGC 6229, NGC 6101, and M5) have RRab stars outside the boundaries matching RRc types. We note that these clusters are not unusually highly reddened, since the mean value for this sample is $\mathrm{E(B-V)} = 0.04$.

We searched for the helium abundance that best matches both RRab and RRc groups. To this end, we defined the IS boundaries using the 68th percentile of the density distribution of RRLs in the color-magnitude diagram. We then varied $Y$ until the difference between the theoretical and empirical IS boundary was minimized. We found that models that minimize this difference for both RRc and RRab stars simultaneously have $Y=0.290$. The density contours for RRab stars are reproduced best by the models with a $Y=0.220$.

We note that double mode (RRd) RRLs fall within the predicted IS boundaries for these models ($Z=0.0003$, $M=0.7$ \(M_\odot\), $Y=0.290$).  Among stars pulsating in the fundamental mode, 80\% are located within the predicted blue and red edges, while 84\% of stars pulsating in the first overtone are within the boundaries set by the models. In other words, $\sim 80\%$ of RRLs are consistent with this single value of $Y$. All outlier RRab stars require significantly lower helium content ($Y = 0.220$), whereas all outlier RRc stars require much higher helium abundance ($Y = 0.357$). 

\begin{figure*}
    \centering
    \includegraphics[scale=0.57]{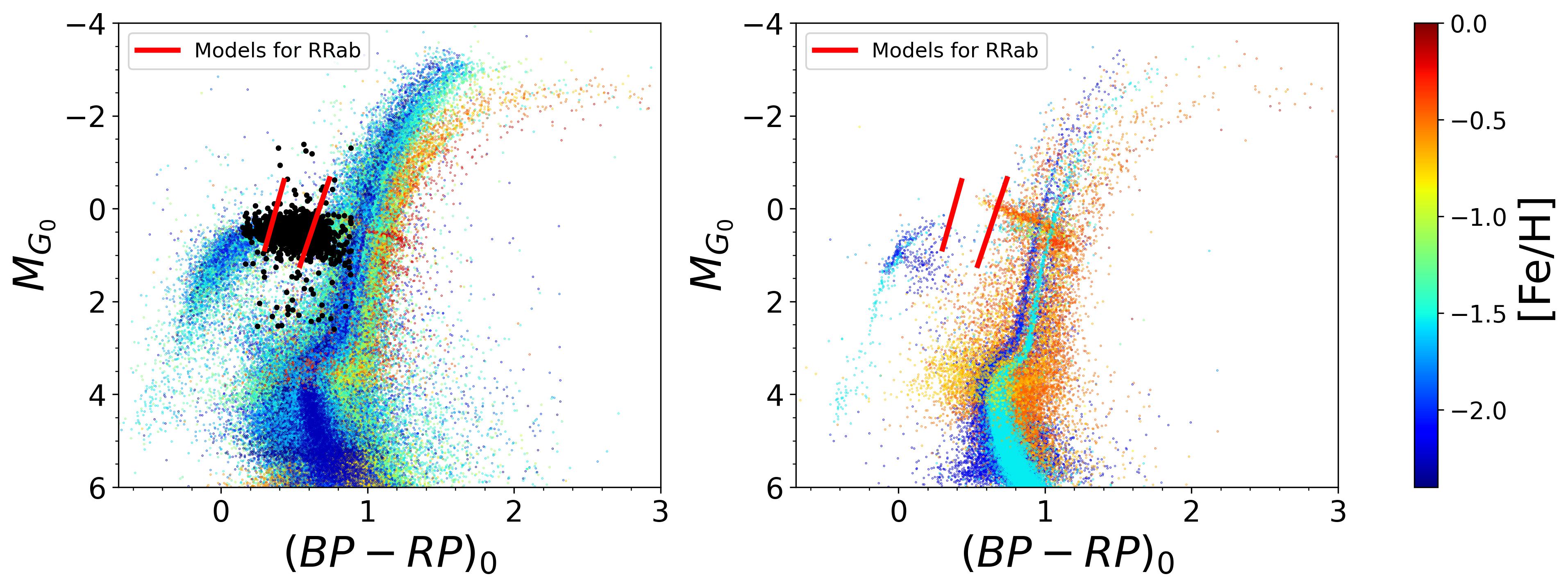}
        \caption{Color-magnitude diagram for the clusters in our sample  $\mathrm{E(B-V)} <1 $, color coded with their mean iron abundance. The plot on the left displays 103 GCs that contain at least one RRL. The right plot shows the same but for 24 clusters without RRLs. The red lines indicate the position of the theoretical blue and red edge of the instability strip for RRab, with $Z=0.0003$, $M=0.7$\(M_\odot\) and $Y = 0.290$, see Sect. \ref{sec:models_is}.}  
    \label{fig:all_rr_lyrae_cleaned}
\end{figure*}

Figure \ref{fig:all_rr_lyrae_cleaned} displays the color-magnitude diagram for all clusters and RRLs in our catalog with $\mathrm{E(B-V)} <1 $. As can be seen in the left panel, some RRLs are outside the HB, this may be related with blending and that we used the photometry from the \texttt{gaia\_source} catalog to make this plot, as not all RRLs have SOS photometry. In the right panel, is interesting to see that clusters lacking RRLs also possess few or no stars on the HB.

Figure \ref{fig:all_rr_lyrae_cleaned} shows some cluster RRLs that fall outside the boundaries of the HB.  Twenty-four percent of them are part of M62. This cluster is located close to the galactic plane in a region where the differential reddening is not negligible, and therefore, a single reddening value is not enough to characterize this cluster.  The rest of the sources outside the HB are characterized by a low posterior probability, with a median value of 0.05. In contrast, the median value for the sources inside the HB is 0.84.

\begin{figure}
    \centering
    \includegraphics[scale=0.34]{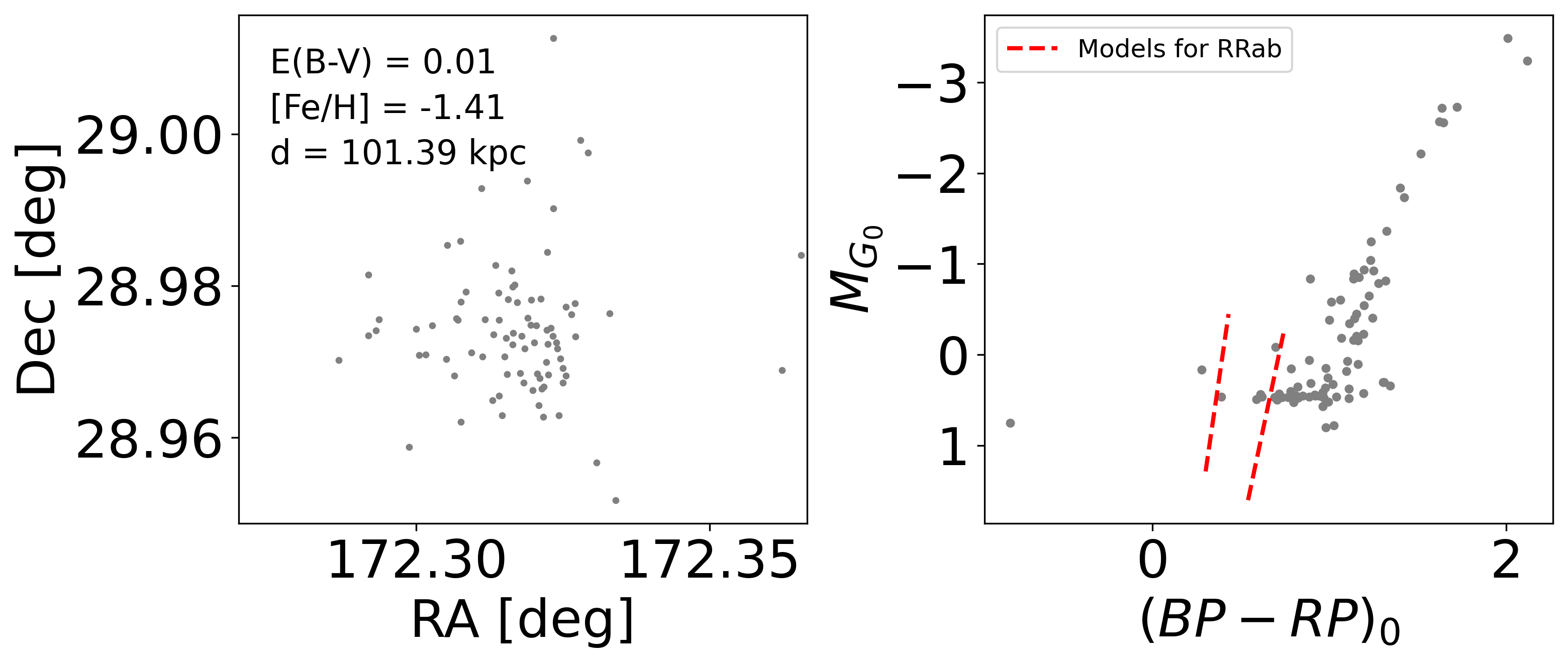}
    \caption{Pal 4, the most distant cluster in our sample with a clearly visible HB, and zero RRL.}
    \label{fig:pal4}
\end{figure}

The cluster Pal~4 seems to be particularly interesting, it is located at $101.39 \pm 2.57$ kpc and the HB of the cluster is visible in Fig. \ref{fig:pal4}, but no RRLs were detected in this study.  NGC~6981 has a similar iron abundance ($\Delta \mathrm{[Fe/H]} = 0.005$) and hosts 43 RRLs, therefore it is likely that Pal~4 hosts a significant number of RRL, but beyond the capabilities of \gaia\ for the detection of RRLs.

\subsection{Purity of the IS}\label{sec:purity_is}
The theoretical models used to describe stellar pulsations, assume that all stars located within the blue and red edge of the IS are pulsating. If this condition is not met, the predictions of stellar populations based on these models would not represent a real population of stars. Interestingly, using ground-based photometry \citet{2018AcA....68..237R} detected one star in NGC~6254~(M10) located in the IS for RRLs that is not variable at the 0.01 mag level. By crossmatching that star with \gaia\ DR3 we confirm that within the limits of \gaia\ that star does not show photometric variability. The dataset of VB21 assigns a membership probability of $\sim 1$ to this star. 

In the case of classical Cepheids in the Large Magellanic Cloud (LMC), it was discovered by \citet{2019MNRAS.489.3285N} that up to $30\%$ of the stars located in the classical instability strip do not show photometric variability at the milimag level. 

We decided to analyze this effect for RRLs using the cluster sample in the Appendix \ref{app:IS}. The results are shown in Fig. \ref{fig:purity_is_color}. It can be observed that the purity of the IS significantly changes as a function of the intrinsic color and absolute magnitude of the stars. It is highest at the center of the IS, gradually decreases toward the axes, approaching zero near the edges. In the following region 
\begin{align}\label{eq:o2}
    0.31 < \mathrm{(BP - RP)}_{0} < 0.67 \quad 0.24 < M_{G_{0}} < 0.92
\end{align}
and within the limits of the method presented in Sect. \ref{sec:HB}, not all GC member stars are detected as RRLs. We found that 25\% of all sources (315 stars) in this region do not exhibit photometric variability. Fifty seven clusters contain at least one non-variable star each inside this region. The median number of non-RRLs in the IS per cluster is four but there are some clusters with an abnormal number of constant stars such as M3 or NGC~6402 that contain 16 and 19 respectively. This stars could be undetected RRLs or interesting new sources. A good way to identify if their photometry is affected by close companions is by using the ipd\_frac\_multi\_peak parameter from \gaia, which provides the fraction of windows for which the Image Parameter Determination (IPD) has identified more than one peak. We do not find evidence of blending in the photometry of the nonvariable sources as the median value of ipd\_frac\_multi\_peak for them is equal to one.  Table \ref{Tab:constant_stars} lists the host cluster, source id, and membership probability computed by VB21 for the non-variable stars. We do not find a clear correlation between the number of non-variable stars and the cluster distance, indicating that this is not an observational bias related with the ability to detect RRLs at large distances. We found that the constant stars are more abundant in clusters with $\mathrm{[Fe/H]}\sim -1.53$.

\begin{table*}{}
  \centering
  \caption{Nonvariable stars located in the IS for RRLs. }
\begin{tabular}{ccc}\toprule  
\gaia\ DR3 source id  &  Host cluster	& VB21 membership probability  \\ 
        \midrule
5771814031481280000  & IC 4499 & 0.99994 \\
6078980613509129472  & Rup 106 & 0.99969 \\
6023588042325345152  & Terzan 3 & 0.99718 \\
...  & ... & ... \\
              \bottomrule
\end{tabular}
\tablefoot{The complete version of this table is available at the CDS. }
    \label{Tab:constant_stars}
\end{table*}

\begin{figure}
    \centering
    \includegraphics[scale= 0.37]{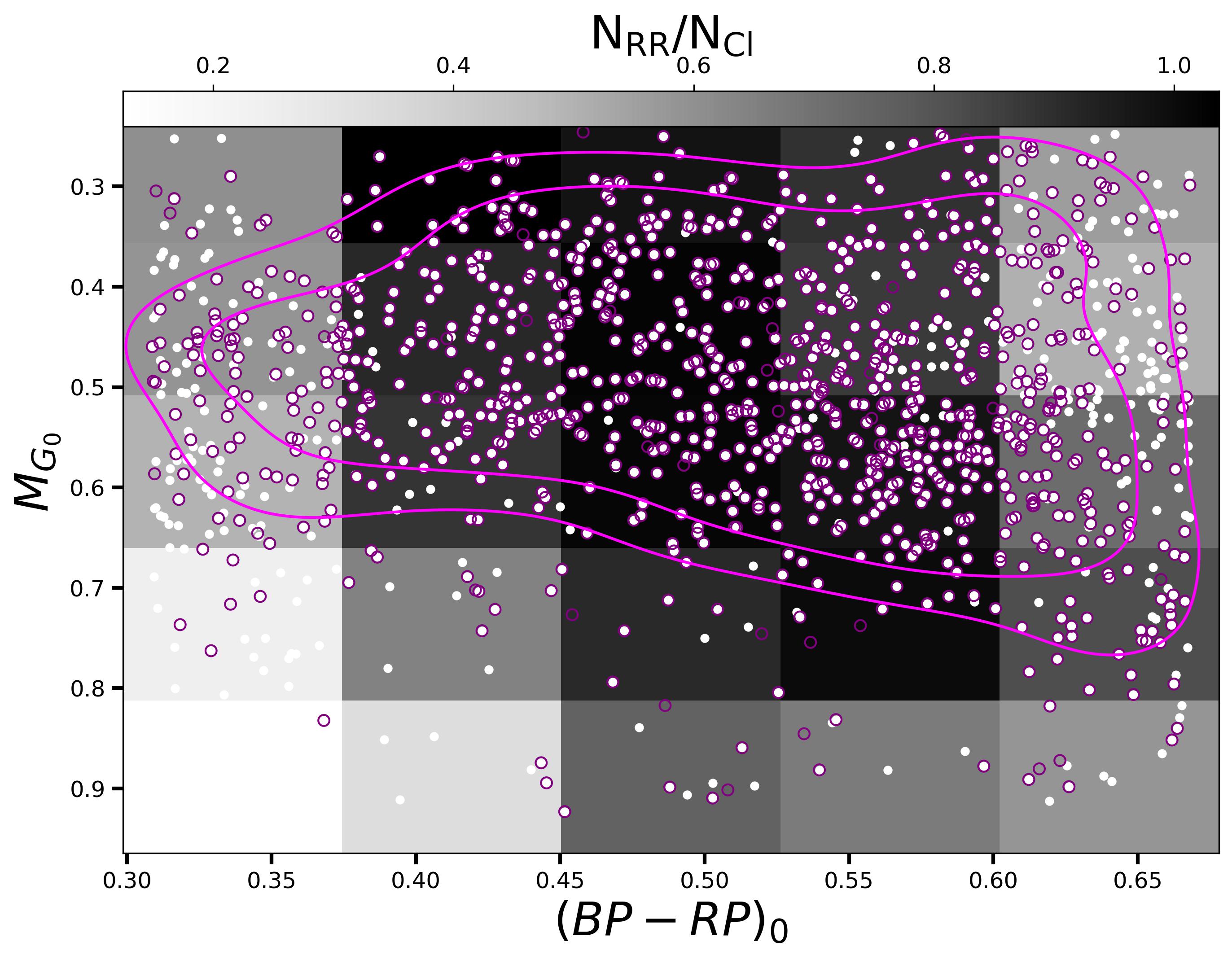}
        \caption{RRLs (pink open circles) in clusters detected in this study and cluster members (white dots) in the HB detected by VB21 with membership probabilities above $50\%$.  The squares indicate the bins in color and absolute magnitude used to estimate the fraction of RRLs over the number of cluster members detected by VB21. Each bin was color coded according to its specific RRL fraction,  the color bar is located at the top of the plot. The purple contours indicate the 68th and 84th percentiles of density of RRLs.}
    \label{fig:purity_is_color}
\end{figure}

\subsection{Oosterhoff dichotomy}\label{sec:Oosterhoff}
The Oosterhoff dichotomy is one of the most well-known conundrums associated with RRLs \citep{1939Obs....62..104O}. The dichotomoy refers to RRLs in clusters being divided into two groups based to their average periods. The first group is known as OoI and has an average period $\left \langle P_{ab} \right \rangle \sim 0.56$ and tends to be metal rich ($\mathrm{[Fe/H]}>-1.5$), while the second group has an average period $\left \langle P_{ab} \right \rangle \sim 0.66$ and is rather metal-poor ($\mathrm{[Fe/H]}<-1.5$). While the Oosterhoff gap refers to the low number of clusters with RRLs in the region $0.58< \left \langle P_{ab} \right \rangle < 0.62 $. This effect has been detected in the Milky Way but not in the Magellanic Clouds \citep{smith2010}. \citet{2019ApJ...882..169F} speculates that the lack of clusters in the gap is associated with the absence of metal-intermediate clusters hosting RRLs.  

\begin{figure*}
    \centering
    \includegraphics[scale=0.5]{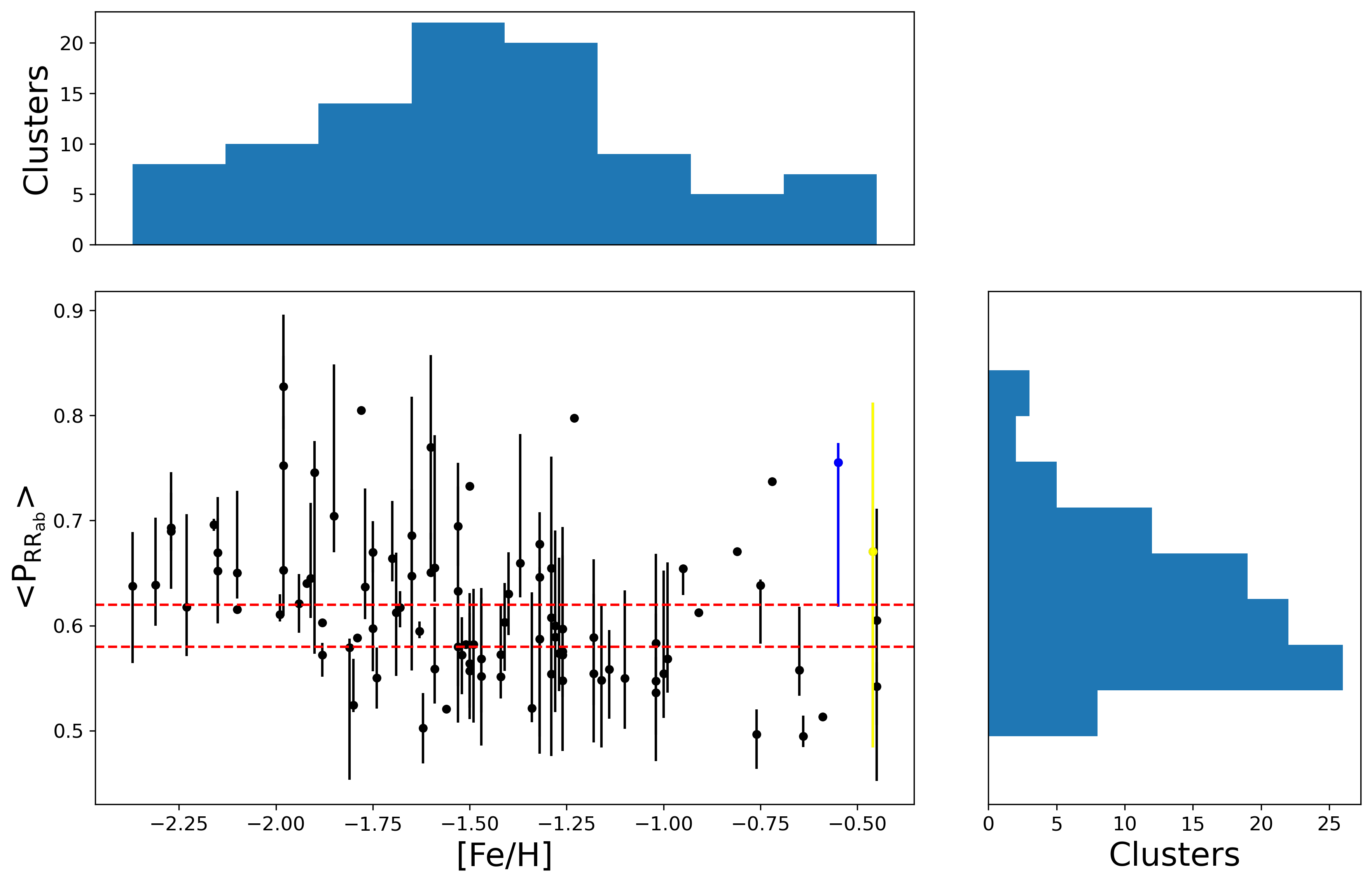}
    \caption{Mean period of the RRLs pulsating in the fundamental mode as a function of the iron abundance of the cluster. The dotted lines indicate the region known as the Oosterhoff gap, we can see that multiple clusters are located in this region. The error bars represent the 16th and 84th percentiles of the distribution of periods.   The blue point represents NGC~6388, and the yellow one corresponds to NGC~6441. Both clusters have been classified   as Oosterhoff type III in the literature  \citep{2003AJ....126.1381P,2022Univ....8..122B}.  }  
    \label{fig:Oosterhoff_gap}
\end{figure*}

We decided to analyze this effect using the RRLs that pulsate in the fundamental mode from \gaia, in our analysis we used the pulsation periods from the SOS analysis. In Fig. \ref{fig:Oosterhoff_gap} it can be observed that there are 22 clusters out of 97 located in the region known as the Oosterhoff gap. Some of those clusters were previously detected by other studies, such as NGC~6626 \citep{2012A&A...543A.148P}, Arp~2 \citep{2019PASP..131e4202P}, NGC~6864 \citep{2005AAS...20712214C}, Rup~106 \citep{2007ApJ...670..332G}, NGC~1851 \citep{jang2014multiple}, NGC~6402 \citep{2022MNRAS.511.1285Y}, and M54 \citep{2016A&A...592A.120F}. Given the large dispersion in the mean value of the periods and the high number of clusters located in the region $0.58< \left \langle P_{ab} \right \rangle < 0.62 $ we consider that is not possible to conclude that this region is a real gap.

To verify if the distribution of $\left \langle P_{ab} \right \rangle$ is described by one or two populations (Oosterhoff types), we fit two models to the observations. The first model is a single Gaussian with three free parameters: mean $\mu$, standard deviation $\sigma$, and amplitude $A$. The second model consists of two Gaussian distributions with three free parameters for each, that is, one Gaussian for each Oosterhoff type. To distinguish which model provides the best  fit to the data, we employ the F-test. The purpose of the F-test is to evaluate how well two distinct models fit the data, while taking into account that these models have different degrees of freedom.

The model with two Gaussians must satisfy the following constraints: First, the mean of the Gaussian distribution assigned to OoI must be smaller than 0.58, otherwise, the mean would fall within the Oosterhoff gap. Similarly, the mean of the distribution assigned to OoII must be greater than 0.62, otherwise the mean would be within the gap. Secondly, the amplitude of the Gaussians obtained from fitting the data should be approximately equal to the number of observed clusters in the period where the peak of the Gaussian is reached. This is important because sometimes there are Gaussians that fit the data but have amplitudes inconsistent with the observations. We chose the difference to be smaller than ten, although the first constraint is sufficient to rule out most models with two Gaussians.

To explore how our results are influenced by the selection of data binning, we repeated our analysis using different bin sizes. We selected the size of the bin with $( P_{max} -  P_{min} )/N_{bins}$, where $P_{max}$ is the maximum average period of the RRLs hosted by a cluster in our sample, $P_{min}$ is the minimum average period and $N_{bins}$ is the number of bins. In total, we used 23 different values of $N_{bins}$, spanning from 7 to 30, which corresponds to bin size ranging from 0.011 to 0.047 days. The two-Gaussian model never passes the F-test and simultaneously meets the established constraints, therefore we conclude that the one-Gaussian model provides a better description of the data. Our results indicate that is not necessary to separate the cluster population into the classical Oosterhoff types.

Twenty-three percent of the clusters shown in Fig. \ref{fig:Oosterhoff_gap} are located within the gap. One could speculate that these clusters have a limited number of RRab, causing significant uncertainties in their mean period. For this reason, we decided to focus only in clusters with more than ten RRab stars. In this case, we found that 19\% of the GCs are within the gap, reaffirming our earlier conclusion that this region is not a gap. 

\begin{figure}
    \centering
    \includegraphics[scale=0.36]{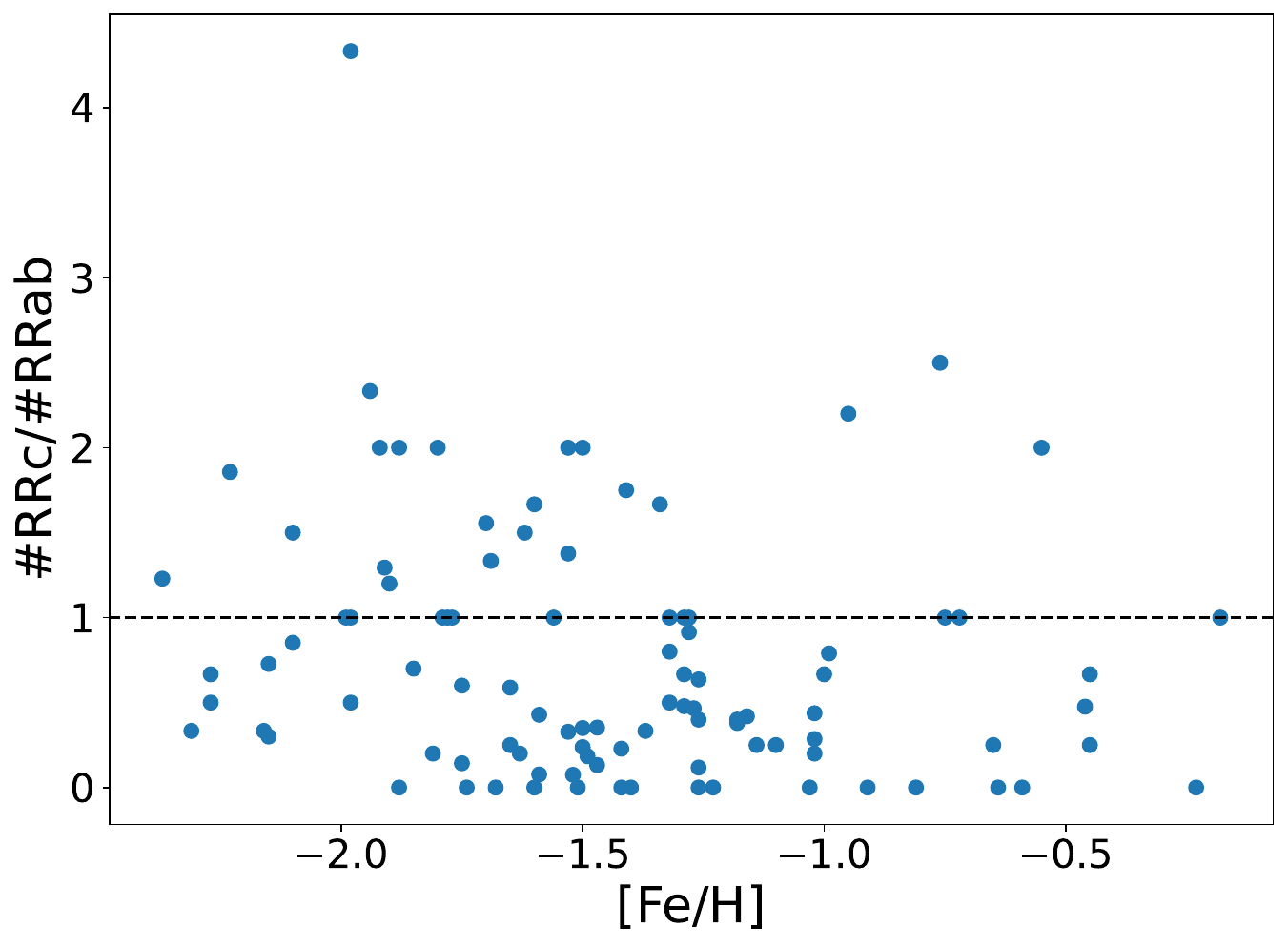}
    \caption{Number of RRLs pulsating in the first overtone divided by the number of the ones pulsating in the fundamental mode. The black dashed line represents unit ratio.} 
    \label{fig:fraction_rr_fe/h}
\end{figure}

Figure \ref{fig:fraction_rr_fe/h} shows the fraction of RRLs pulsating in the fundamental mode over those pulsating in the first overtone as a function of the iron abundance. It can be observed that RRc stars are more abundant than RRab stars in metal-poor clusters, and this trend is reversed for metal-rich clusters. This is expected because RRc stars are bluer than RRab stars,  and metal-poor clusters are characterized by having bluer HBs than metal-rich clusters.

\begin{figure*}
    \centering
    \includegraphics[scale=0.4]{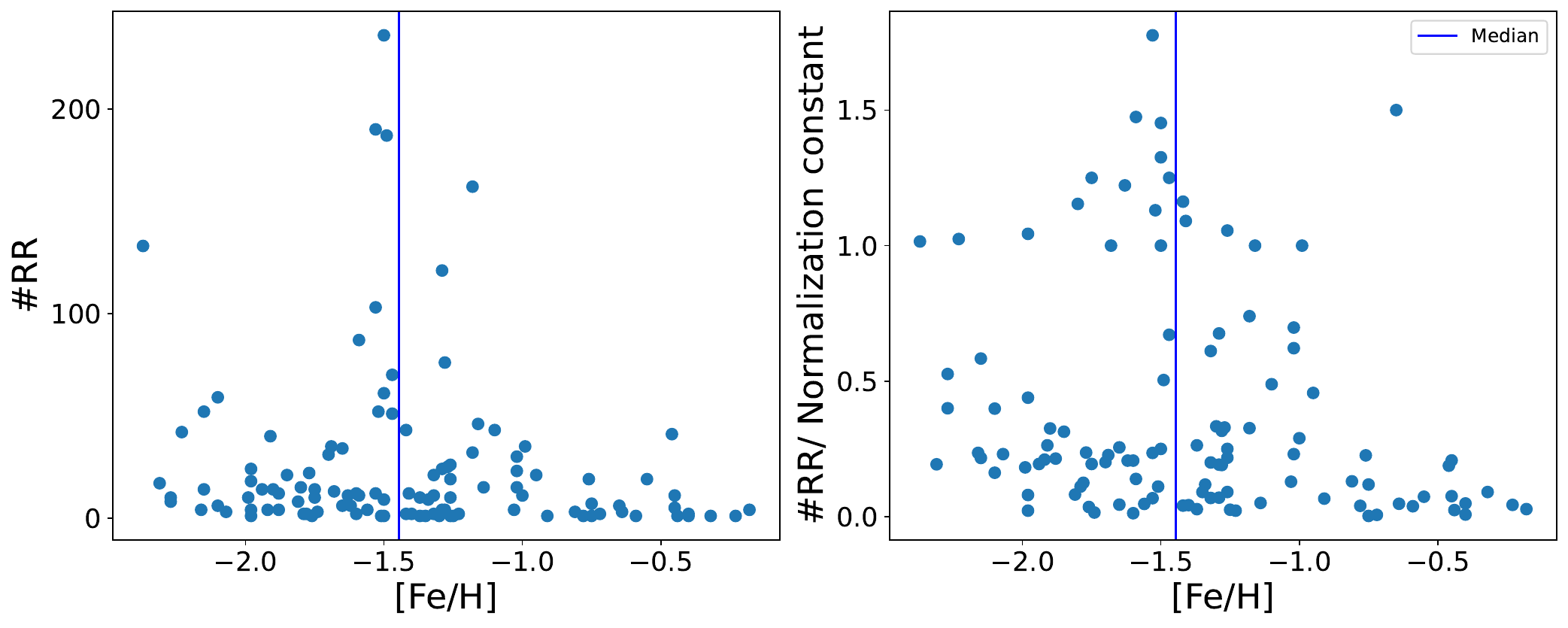}
    \caption{Number of RRLs in clusters. The panel illustrates the number of RRLs within each cluster as a function of the iron abundance. The right panel presents the same data, but the number of RRLs is divided by a normalization constant that is calculated individually for each cluster (refer to Sect. \ref{sec:Oosterhoff} for more details).    }
    \label{fig:number_rr}
\end{figure*}

The left panel of Fig. \ref{fig:number_rr} shows that the MW clusters with the most RRLs have  $\mathrm{[Fe/H]} = -1.44$. However, the number of RRLs can also depend on observational selection effects, such as the number of stars in the clusters and their distance. The right panel of Fig. \ref{fig:number_rr} thus illustrates the number of RRLs divided by the number of stars located on the red giant branch, $R_{\mathrm{RGB}}^{\mathrm{RRL}}$, within one and two magnitudes above the apparent magnitude of RRLs in each cluster. We chose this specific range to estimate the normalization constant because the number of stars in the HB varies significantly among different clusters. Even after introducing the normalization, RRLs remain particularly common in clusters with iron abundance $\mathrm{[Fe/H]}\approx -1.5$. However, we identify a surprising dichotomy among clusters according to the ratio $R_{\mathrm{RGB}}^{\mathrm{RRL}}$in this ratio. The majority of clusters prefers low $R_{\mathrm{RGB}}^{\mathrm{RRL}} \lesssim 0.2$, that is, there typically at least four red giants between one and two magnitudes brighter than RRLs. However, $R_{\mathrm{RGB}}^{\mathrm{RRL}}$ can be significantly larger, reaching up to $> 1.5$, implying more RRL than RGB stars in this magnitude bin. Curiously, there are many clusters with an even ratio, $R_{\mathrm{RGB}}^{\mathrm{RRL}} = 1$. At present, it is not clear what causes this feature, which could be related to age or observational selection effects.

\section{Summary and conclusions}\label{sec:summary}
This is the first paper of a series dedicated to study variable stars in GCs. By making use of \gaia's astrometry and photometry, our catalog improves the confidence of membership of multiple RRLs to GCs.  Despite the dense stellar environments within GCs, our analysis demonstrated the robustness of the classification methods of the \gaia\ collaboration for RRLs, as most of the identified RRLs are located in the HB.

Currently, our catalog is the most comprehensive one obtained with a homogeneous set of astrometry and photometry.  However, the limitations of \gaia's angular resolution and its poorer astrometry at the center of GCs pose challenges to our analysis, potentially resulting in an incomplete sample within those regions.  It is worth noting that studies dedicated to specific clusters may be more complete than our catalog. For example, \citet{2023RMxAA..59....3A} has found that there are 18 RRLs in Palomar~2. Whereas we only detected two, some of the stars detected by \citep{2023RMxAA..59....3A} do not appear in our analysis because they are outside the range where the L21 corrections are defined, and the rest are removed due to a low likelihood or prior. The clusters 2MASS-GC01, 2MASS-GC02, GLIMPSE01, and GLIMPSE02 do not appear in our analysis because they are not listed in the VB21 catalog of GCs,  as they are located in highly extincted regions, thus for them, studies in the infrared will be better suited to study their population of RRLs. 

We show that more than 80\% of RRLs are located within the instability strip boundaries predicted by the MESA models with $Z=0.0003$, $M=0.7$ \(M_\odot\), and $Y=0.290$. However, the models that best describe the population of RRab stars feature a lower helium content (Y = 0.220), and the models that best describe the population of RRc stars require a higher helium abundance, Y = 0.357.

The observation that a significant portion ($25\%$) of the stars located in the observational instability strip do not exhibit pulsations challenges our understanding of stellar pulsations models. Further research is needed to explore the reasons behind this behavior, potentially leading to advancements in stellar evolution theories. The finding that the Oosterhoff dichotomy does not exist in our sample suggests that the traditional classification of RRLs based on this dichotomy is not necessary.

In our dataset, there are clusters such as NGC~2419 or Pal~3 which are located at $88.47 \pm 2.40$~kpc and $94.84 \pm 3.23$~kpc and contain a significant number of RRLs (52 and 11 respectively) with an average magnitude $G\approx 20.2$. The faintest stars that \gaia\ can detect are of magnitude $G \approx 21$, and considering that the amplitude of this type of stars in the $V$ band can reach up to $1.5$ magnitudes, it means that we are at the limit of Gaia's detection capabilities. The next paper of the series will focus on the calibration of the period-luminosity relations for multiple types of pulsating stars.

\begin{acknowledgements}
We are thankful for useful discussions with Laurent Eyer.

MC, RIA, and HN acknowledge support from the European Research Council (ERC) under the European Union's Horizon 2020 research and innovation programme (Grant Agreement No. 947660). RIA is funded by the SNSF through a Swiss National Science Foundation Eccellenza Professorial Fellowship (award PCEFP2\_194638). 

This research was supported by the Munich Institute for Astro-, Particle and BioPhysics (MIAPbP) which is funded by the Deutsche Forschungsgemeinschaft (DFG, German Research Foundation) under Germany's Excellence Strategy – EXC-2094 – 390783311.

This work has made use of data from the European Space Agency (ESA) mission
{\it Gaia} (\url{https://www.cosmos.esa.int/gaia}), processed by the {\it Gaia}
Data Processing and Analysis Consortium (DPAC,
\url{https://www.cosmos.esa.int/web/gaia/dpac/consortium}). Funding for the DPAC
has been provided by national institutions, in particular the institutions
participating in the {\it Gaia} Multilateral Agreement. 

This research has made use of NASA's Astrophysics Data System; the SIMBAD database and the VizieR catalog access tool\footnote{\url{http://cdsweb.u-strasbg.fr/}} provided by CDS, Strasbourg; Astropy\footnote{\url{http://www.astropy.org}}, a community-developed core Python package for Astronomy \citep{astropy:2013, astropy:2018}; TOPCAT\footnote{\url{http://www.star.bristol.ac.uk/~mbt/topcat/}} \citep{2005ASPC..347...29T}.

\end{acknowledgements}

\bibliographystyle{aa}
\bibliography{RR}

\begin{appendix}

\section{Cluster parameters}\label{sec:parameters}
Cluster central positions in the sky were computed as the mean RA and DEC of all sources from the VB21 catalog with membership probabilities above $50\%$ and without using astrometric or photometric quality cuts. The position angles and ellipticities, are obtained with the method explained in Sect. \ref{sec:clusters} and the uncertainties are estimated using 1000 bootstrap resamples. The parameters are recalculated for each resampling and the standard deviation of the resulting distribution is taken as the uncertainty in those parameters. 

For the rest of the cluster parameters, we used all five and six parameter solutions from \gaia\ within the limits of the astrometric and photometric quality cuts suggested by VB21 with membership probabilities above $90\%$. To guarantee accurate corrections of the parallax offset, we further limited our analysis to the sources that fall within the magnitude and color range specified by the L21 corrections, all our constraints are summarized in Table \ref{tab:constraints_cluster members}.

To compute the cluster parallaxes, we first compute parallax offsets for each cluster member following L21. In turn, we compute the cluster parallax as the weighted mean of the cluster members. The cluster's parallax uncertainty comprises two terms, the angular covariance and the statistical error following \citet{apellaniz2021validation}.  Proper motions are computed as the mean of all cluster members and the uncertainties are the standard deviation on the mean. Given the low number of cluster members with radial velocity measurements, we report the median and the standard error on the median. Table \ref{tab:cluster_parameters} contains a list of the cluster parameters.

We compared our cluster parameters with the ones estimated by VB21. In general, all parameters argreed with VB21 to within the respective uncertainties. Small differences among our and VB21's cluster parallaxes are explained by the use of different quality cuts. However, we did not find any indication of values being biased in a particular direction. The difference in the proper motion uncertainties arises because we use the standard error in the mean while they use the weighted error.

\begin{table*}{}
 \centering
\caption{Astrometric and photometric constraints that were applied to determine the cluster parameters. \label{tab:constraints_cluster members}}
\begin{tabular}{ll}\toprule
\multicolumn{1}{c}{Astrometric constraints} &  \multirow{1}{*}{Photometric constraints}   \\
 \cmidrule(lr){1-1}\cmidrule(lr){2-2} 
ruwe $<1.15$     &   $ 13 < G < 21  $ \\ 
astrometric\_excess\_noise $< 2 $   &  ipd\_gof\_harmonic\_amplitude $ < \exp\left [ 0.18 (G - 33) \right ]$   \\
duplicated sources are removed &  ipd\_frac\_multi\_peak $ < 2$   \\ 
1.24 < pseudocolor< 1.72  & $1.1 <$ nu\_eff\_used\_in\_astrometry $< 1.9$ \\ 
  & visibility\_periods\_used $ > 10 $ \\ 
  & $C^{*}<3 \sigma_{C^{*}}(G)$ \citep{2021AA...649A...3R} \\ 
\bottomrule
\end{tabular}
\end{table*}

\begingroup
\setlength{\tabcolsep}{6.pt} 
\begin{table*}{}
  \centering
  \ra{0.7}
  \setlength{\tabcolsep}{2.pt} 
  \caption{Cluster parameters.}
\begin{tabular}{lcccccccccccc}\toprule  
Cluster & N  & $\varpi$ & $\mu_{\alpha}^{*}$ & $\mu_{\delta}$ & RV   & $\mathrm{E(B-V)}$ & N PCA  & $\theta$ &  $a_{c}$  & $a_{lim} $ & $\epsilon \times 100$ &  N RRL  \\ 
        &    & ($\mu$as)   & (mas/yr)        & (mas/yr)       & (km/s) &       &      &      & ($^{\circ}$) &    \\ 
 \midrule
NGC 5139 & 69415 & $190 \pm 9$  & $-3.245 \pm 0.620$ & $-6.787 \pm 0.590$ & $234.5 \pm 0.9$ & 0.12 & 147516  &  $164	\pm 1$ &  0.326 & 0.693 & $6.4\pm 0.2$ & 190 \\
NGC 104  & 53656  & $226 \pm 10$ & $5.272 \pm 0.531$  & $-2.547\pm 0.562$ & $-16.8 \pm 0.4$ & 0.04 &102838  &  $47 \pm 2$   & 0.325 & 0.684 & $5.9\pm 0.4$  & 2 \\
NGC 6752 & 20431  & $254 \pm 10$ & $-3.161 \pm 0.458$ & $-4.042	\pm 0.462$ & $-26.7	\pm 1.0$ & 0.04 &  38979 & $156 \pm 12$ & 0.209 &  0.440 & $2.3\pm 0.6$  & 0 \\
... &  ...  &  ... &  ... &  ... & ...  & ... & ... & ... & ... & ... & ...\\
              \bottomrule
\end{tabular}
\tablefoot{The complete version of this table is available at the CDS. Cluster name, number of cluster members used for the determination of the astrometric parameters, proper motion in right ascension and declination together with the error on the mean, radial velocity and error on the median. The main source of color excess measurements is the \citet{harris2010new} catalog, for FSR~1758 we used the value provided by \citet{2021AA...652A.158R}. Number of cluster members used for the determination of $a_{c}$, $a_{lim}$ (see Sect. \ref{sec:Prior}), position angle and number of RRLs in each cluster. }
    \label{tab:cluster_parameters}
\end{table*}
\endgroup

\section{Purity of the IS}\label{app:IS}

The following defines the sample of clusters used to analyze the population of RRLs within clusters in Sects. \ref{sec:M_analysis} and \ref{sec:purity_is}.  It is crucial to guarantee that all stars under consideration during the analysis of IS purity are genuinely located within it. Incorrect extinction values can affect the position of stars in the color-magnitude diagram, and therefore we restrict our analysis to clusters with $\mathrm{E(B-V)}<1.0$ (84th percentile of the distribution of $\mathrm{E(B-V)}$).   For cluster members and RRLs, we required a minimum of ten epochs in $G$, $G_{BP}$, and $G_{RP}$. To avoid sources affected by photometric blending, we excluded sources from our analysis for which more than 10$\%$ of the transits were contaminated by the light from a nearby star (ipd\_frac\_multi\_peak <10) and we select sources with ruwe$<1.4$. To guarantee an accurate determination of the cluster parameters, we select clusters with at least 300 members (median number  of members in a cluster after quality cuts). Finally, to avoid potential contamination of background or foreground stars we restrict our sample to clusters at distances smaller than 23 kpc (84th percentile of the distribution of distances). In total, from the 170 initial clusters, only 75 in the VB21 meet all constrains. The quality cuts are summarized in Table \ref{tab:cuts_IS}.

\begin{table*}{}
 \centering
\caption{Astrometric and photometric constraints applied to the analysis of the purity of the instability strip.\label{tab:cuts_IS}}
\begin{tabular}{cccc}\toprule
\multicolumn{1}{c}{Astrometric constraints} &  \multirow{1}{*}{Photometric constraints} &  \multirow{1}{*}{Constrains for clusters}   \\
 \cmidrule(lr){1-1}\cmidrule(lr){2-2} \cmidrule(lr){3-3} 
ruwe $<1.4$     &   ipd\_frac\_multi\_peak$< 10$ &      $\mathrm{E(B-V)} <1$& \\ 
& num\_clean\_epochs\_g, bp and rp $ > 10 $ & Number of cluster members $>300$ \\
&                                           & distance $<23 $ kpc      \\
\bottomrule
\end{tabular}
\end{table*}

\section{Newly detected RR Lyrae}\label{app:New_rr}
Table \ref{tab:new_rr} provides the host cluster, \gaia\ DR3 source id, the classification determined by \gaia\ from \texttt{vari\_classifier\_result} and the additional references for the detected RRLs in Sect. \ref{sec:HB}.

\begin{table*}[!htp]
    \centering
    \caption{RR Lyrae that were detected using the uncertainties in the \gaia\ photometry.}
    \begin{tabular}{@{}lcccl}\toprule
         Cluster  &  \gaia\ DR3 Source id  & Classification from \gaia\ & Previously detected & Reference \\ \midrule
          NGC 6093   & 6050423066717872128 & ECL & Yes  & \cite{2013AcA....63...91K} \\ 
          NGC 6093   & 6050422173364568576 & ECL & Yes  &  \cite{2013AcA....63...91K}\\  
          NGC 5286   & 6069336998880602240 & ECL & Yes  &  \cite{2022arXiv220606278C}\\ 
          NGC 5286   & 6069383594988480768 &     & Yes  & \cite{2010AJ....139..357Z} \\ 
          NGC 6402   & 4368940930792772992 &  ECL  & Yes  & \cite{2018AJ....155..116C}  \\
          NGC 6402   & 4368940789054349184 &     & Yes  & \cite{2018AJ....155..116C} \\         
          NGC 6205  & 1328058970889601408 & ECL & Yes  & \cite{2019MNRAS.486.2791D} \\ 
          NGC 5897 & 6252666475314697600 & ECL  & Yes &  \cite{2001AJ....122.1464C} \\ 
          NGC 6656  & 4077589384806381952 &     & Yes  & \cite{2013AJ....146..119K}\\     
          NGC 6266  & 6029364769043583104 & ECL & Yes  & \cite{2010AJ....140.1766C} \\      
          NGC 4833   & 5843798787193074944 & ECL & Yes  & C17 \\
          NGC 5286   & 6069385003736234240 &     & No  & \\ 
          NGC 3201   & 5413533670752159232 &   & No  & \\   
          NGC 6864   & 6853720936203441792 &     & No  & \\ 
         \bottomrule
    \end{tabular}
    \tablefoot{The RR Lyrae with source id 6069336998880602240, is not included in  \texttt{vari\_rrlyrae}, however it was detected and characterized by the \gaia\ collaboration \citep{2022arXiv220606278C}.   }
    \label{tab:new_rr}
\end{table*}

\section{SOS sources with unrealistic photometric uncertainties.}\label{app:sources_sos}

During the creation of this catalog, we detected some issues in the parameters of 63 stars present in \texttt{vari\_rrlyrae}, they can be found in Table \ref{Tab:sources_sos}.  Some of them have uncertainties of the order of $10^{-14}$ mag in BP (or RP), but this precision is beyond the capabilities of \gaia.
Those sources are characterized by less than 10 epochs in BP (or RP). Additionally, we detected some sources with negative apparent magnitudes and this should not be observable by \gaia. We would like to emphasize that the number of stars exhibiting these characteristics is minimal and does not reflect the overall quality of the SOS analysis.

\begin{table*}{}
  \centering
  \ra{0.7}
  \addtolength{\tabcolsep}{-2.pt} 
  \caption{Sources in the Specific Object Study for RRLs with unrealistic photometric uncertainties. }
\begin{tabular}{cccccc}\toprule  
source\_id  & int\_average\_bp\_error	& num\_clean\_epochs\_bp & int\_average\_rp\_error	& num\_clean\_epochs\_rp \\  
          &   mag           &              &       mag                   &               &                        \\ 
        \midrule
4295843576778048384  & $20.1 \pm 7\times 10^{-15}$ & 6 & $20.1	\pm 7\times10^{-15}$ & 6\\
4654634767984918528  &  $19.2	\pm 4\times 10^{-15}$	& 9& $19.2\pm	4\times 10^{-15}$ & 9 \\
...  & ... & ...  & ... &\\

              \bottomrule
\end{tabular}
\tablefoot{The complete version of this table is available at the CDS. }
    \label{Tab:sources_sos}
\end{table*}

\end{appendix}

\end{document}